\documentclass[12pt,a4paper]{article}

\usepackage[includeheadfoot,
            marginratio={1:1,2:3}, 
            width=412pt, 
            height=688pt,]{geometry}

\usepackage{amsmath,mathrsfs}
\usepackage{amsfonts}
\usepackage{amssymb}
\usepackage{stmaryrd}
\usepackage{graphicx}
\DeclareMathAlphabet{\mathpzc}{OT1}{pzc}{m}{it}

\usepackage{empheq}
\usepackage{paralist}

\newcommand{\nc}{\newcommand}
\nc{\lb}{\llbracket}
\nc{\rb}{\rrbracket}
\nc{\gl}{\llbracket}
\nc{\gr}{\rrbracket}

\newcommand{\eq}[1]{\begin{equation}
                     \begin{split} #1 \end{split}
                     \end{equation}}

\allowdisplaybreaks[2]
\numberwithin{equation}{section}
\begin{document}
\begin{titlepage}

%\begin{flushright}

\normalsize

\begin{center}

\vspace{1cm}

\baselineskip=24pt

{\Large\bf Non-commutative deformation of Chern-Simons theory}

\baselineskip=14pt

\vspace{1cm}

{\bf Vladislav G. Kupriyanov}
\\[5mm]
\noindent {\it CMCC-Universidade Federal do ABC, Santo Andr\'e, SP, 
Brazil}\\
{\it Tomsk State University, Tomsk, Russia}
\\
Email: \ {\tt
    vladislav.kupriyanov@gmail.com}
\\[30mm]

\end{center}

\begin{abstract}
\baselineskip=12pt
\noindent
The problem of the consistent definition of gauge theories living on the non-commutative (NC) spaces with a non-constant NC parameter $\Theta(x)$ is discussed. Working in the L$_\infty$ formalism we specify the undeformed theory, $3$d abelian Chern-Simons, by setting the initial $\ell_1$ brackets. The deformation is introduced by assigning the star commutator to the $\ell_2$ bracket. For this initial set up we construct the corresponding L$_\infty$ structure which defines both the NC deformation of the abelian gauge transformations and the field equations covariant under these transformations. To compensate the violation of the Leibniz rule one needs the higher brackets which are proportional to the derivatives of $\Theta$. Proceeding in the slowly varying field approximation when the star commutator is approximated by the Poisson bracket we derive the recurrence relations for the definition of these brackets for arbitrary $\Theta$. For the particular case of $su(2)$-like NC space we obtain an explicit all orders formulas for both NC gauge transformations and NC deformation of Chern-Simons equations. The latter are non-Lagrangian and are satisfied if the NC field strength vanishes everywhere.

\end{abstract}

\end{titlepage}

\setcounter{page}{2}

\newpage

{\baselineskip=12pt
\tableofcontents
}

\bigskip

\renewcommand{\thefootnote}{\arabic{footnote}}
\setcounter{footnote}{0}

\section{Introduction}

In the standard approach to the definition of gauge theory one needs the notion of the covariant derivative, $\mathcal{D}_a=\partial_a-iA_a$, as a generalization of the partial derivative $\partial_a$ which transforms covariantly, $\mathcal{D}_a\to e^{if(x)}\mathcal{D}_a$, under the gauge transformations $\delta_f A_a=\partial_a f$. This notion is based on the Leibniz rule. The non-commutativity is a fundamental feature of the space-time which manifests itself at very short distances \cite{Doplicher, SW}. It can be introduced in the theory through the star product, defined in the theory of deformation quantization \cite{BFFLS} as,
\eq{f\star g=f\cdot g+\frac{i}{2}\,\Theta^{ab}(x)\,\partial_a f\partial_b g+\dots\,,} 
where $\Theta^{ab}(x)$ is the non-commutativity parameter depending on the specific physical model. In some cases, like the open string dynamics in the constant $B$-field \cite{SW}, the non-commutativity parameter can be constant, however in general it is a function of coordinates. The coordinate dependence of $\Theta$, in general, leads to the violation of the Leibniz rule,
\eq{
\partial_c(f\star g)=(\partial_c f)\star g+f\star(\partial_c g)+\frac{i}{2}\,\partial_c\Theta^{ab}(x)\,\partial_a f\partial_b g+\dots\,,} 
and makes it impossible to follow the standard path for the formulation of NC gauge theory. Let us note that in some particular cases, like the NC gauge theory on D-branes in non-geometric backgrounds \cite{HS} the type of non-commutativity is compatible with the Leibniz rule, so the standard reasoning can be used for the definition of the NC field strength. Because of the non-geometry one has to shift the field strength tensor by a closed two-form on the D-brane worldvolume to construct the NC Yang-Mills action.

The problem with the violation of the Leibniz rule can be taken under control if, e.g., instead of the partial derivative $\partial_a$ one takes the inner one defined through the star commutator, $D_a= i[\,\cdot\,,x_a]_\star$, as was done in the approach of covariant coordinates \cite{Wess}. This however may lead to a problem with the correct commutative limit. Another possibility discussed in the literature consists in using the deformed Leibniz rule constructed with the help of the twist element of a Hopf algebra \cite{Dimitrijevic, Vassilevich}. Here we mention that the twist element is known for the very few examples of NC spaces \cite{Szabo}.

In recent work \cite{BBKL} in collaboration with Ralph Blumenhagen, Ilka Brunner and Dieter L\"ust we have formulated the L$_\infty$-bootstrap approach to the construction of non-commutative gauge theories. On the one hand, in the physical literature L$_\infty$ structures were introduced for description of gauge theories \cite{Zwiebach}, see also \cite{HZ,Saemann} for more details and recent references. On the classical level such an L$_\infty$ structure contains all necessary information about the theory including the gauge symmetry, the field equations and the Noether identities. On the other hand, L$_\infty$ algebras (also known as strong homotopy Lie algebras) \cite{Stasheff1,Stasheff2} provides a framework for dealing with the deformation since the Jacobi identities are required to hold only up to a total derivative or a higher coherent homotopy. We note in particular that the proof of the key result in deformation quantization, the Formality Theorem, is based on the concept of L$_\infty$ algebras \cite{Kontsevich}.

The main idea of the L$_\infty$ bootstrap approach consists in two steps. The first one is to represent the original undeformed gauge theory, like the Chern-Simons or the Yang-Mills, as well as the deformation introduced through the star commutator as a part of a new L$_\infty$ algebra specifying the initial brackets $\ell_1$, $\ell_2$, etc. Then solving the L$_\infty$ relations, $\mathcal{J}_n=0$, one determines the missing brackets $\ell_n$ and completes the L$_\infty$ algebra which governs the NC deformation of the gauge transformations and the equations of motion. In \cite{BBKL} we found the expressions for the gauge transformations and the field equations up to the order $\mathcal{O}\left(\Theta^2\right)$ in the non-commutativity parameter. However the calculations were extremely involved and it was not clear whether the procedure can be extended to the higher or potentially all orders in the deformation parameter.

The purpose of the current work is to develop the ideas and tools proposed in \cite{BBKL} for the construction of the solution for the L$_\infty$ bootstrap program and to apply them to the explicit example, the non-commutative deformation of the abelian Chern-Simons theory.  The key observation made in this paper is that in each given order $n$ there is a set of \emph{consistency conditions} for solvability of the L$_\infty$ bootstrap equations, $ {\cal J}_n=0$. These conditions are satisfied as a consequence of the previously solved equations, ${\cal J}_m=0$, with $m<n$. We use this observation to express the brackets $\ell_n$ in terms of those which have already been found. Note that the solutions $\ell_n$ of the equations, ${\cal J}_n=0$, are not unique and often can be chosen to be zero. Aiming to provide explicit calculations we work in the slowly varying field approximation when the higher derivative terms in the star commutator are discarded and it is approximated by the Poisson bracket. We set,
\eq{\label{ell2}
\ell_2(f,g)=-\{f,g\}=-\Theta^{ab}(x)\,\partial_a f\,\partial_b g\,,
}
where $\Theta^{ab}(x)$ is any suitably symmetric function satisfying the Jacobi identity, 
\eq{\label{jactheta}
\Theta^{al}\,\partial_l\,\Theta^{bc}+\Theta^{cl}\,\partial_l\,\Theta^{ab}+\Theta^{bl}\,\partial_l\,\Theta^{ca}=0\,.
}

The paper is organized as follows. We start with a brief review of basic facts from L$_\infty$ algebras in the Sec. 2. In the Sec. 3 we construct the NC deformation of the abelian gauge transformation,
\eq{ \label{ih2}
\delta_f A_a=\partial_a f+\{A_a, f\}+ \Gamma^k_a(A)\,\partial_kf\,,
}
which satisfies the gauge closure condition, $[\delta_{f},\delta_g] A_a =\delta_{\{f,g\} }A_a$. Then, in the Sec. 4 we derive an expression for the field equations,
\eq{
\label{ie2}
   {\mathcal  F}^a:=P^{abc}\left(A\right)\,\partial_b A_c+R^{abc}\left(A\right)\,\left\{A_b,A_c\right\}=0\,,
}
which are covariant under the transformation (\ref{ih2}), i.e., $ \delta_{f}  {\mathpzc F}^a = \{{\mathpzc F}^a,f\}\,,$ and reproduce in the commutative limit, $\Theta\to 0$, the standard abelian Chern-Simons equations, $\varepsilon^{abc}\,\partial_bA_c=0$.
For arbitrary NC parameter $\Theta^{ij}(x)$ we provide the recurrence relations for the construction of the functions $\Gamma^k_a(A)$, $P^{abc}(A)$ and $R^{abc}(A)$ at any order in $\Theta$. For deformation parameter, which is linear in coordinates, in the Sec. 5 we obtain an explicit all orders expressions for both NC gauge transformations and NC field equations. We define the NC field strength and show that just like in the commutative case the NC Chern-Simons equations are equivalent to the requirement that the NC field strength should vanish everywhere.

\section{Basic facts from L$_\infty$-algebras}

For the convenience of the reader in this Section we will briefly review the basic facts form the theory of L$_\infty$-algebras and its relation to the gauge theories. We start with a formal definition. In fact, L$_\infty$-algebras are generalized Lie algebras where one has not only a two-bracket but more general multilinear $n$-brackets with $n$ inputs
\eq{
\ell_n: \qquad \quad X^{\otimes n} &\rightarrow X \\
x_1, \dots , x_n &\mapsto \ell_n(x_1, \dots , x_n) \, , 
}
defined on a graded vector space $X = \bigoplus_m X_m$, where $m\in \mathbb{Z}$, denotes the grading of the corresponding subspace. Each element $x\in X$, has its own degree, meaning that if ${\rm deg}(x)=p$, this element belongs to the subspace $X_p$.
These brackets are graded anti-symmetric,
\eq{ 
\label{permuting}
\ell_n (\dots, x_1,x_2, \dots) = (-1)^{1+ {\rm deg}(x_1) {\rm deg}( x_2)} \, \ell_n (\dots, x_2,x_1, \dots )\,.
}
The degree of $\ell_n$ is,
\eq{\label{degree}
      p={\rm deg}\big( \, \ell_n(x_1,\ldots,x_n)\, \big)=n-2+\sum_{i=1}^n  {\rm deg}(x_i)\,.
}
 The set of higher brackets  $\ell_n$ define an L$_\infty$ algebra if they satisfy the
 infinitely many relations
\eq{
\label{linftyrels}
{\cal J}_n(x_1,\ldots, x_n):=\sum_{i + j = n + 1 } &(-1)^{i(j-1)}
\sum_\sigma  (-1)^\sigma
\, \chi (\sigma;x) \; \\
 &\ell_j \big( \;
\ell_i (x_{\sigma(1)}\; , \dots , x_{\sigma(i)} )\, , x_{\sigma(i+1)} , \dots ,
x_{\sigma(n)} \, \big) = 0 \, .
}
The permutations are restricted to the ones with
\eq{ 
\label{restrictiononpermutation}
\sigma(1) < \cdots < \sigma(i) , \qquad \sigma(i+1) < \cdots < \sigma(n)\,,
}
and the sign $\chi(\sigma; x) = \pm 1$ can be read off from \eqref{permuting}. 

The framework of L$_\infty$ algebras is quite flexible
and it has been suggested  that every classical perturbative gauge
theory
including its dynamics, is organized by an underlying L$_\infty$
structure \cite{HZ}. To see this, let us assume that the field theory has a standard
type gauge structure, meaning that the variations of the fields can be
organized unambiguously into a sum of terms each of a definite power
in the fields. We take the graded space,
$X=X_{0}\oplus X_{-1}\oplus X_{-2}$, and all others trivial. The
assignment is as follows. $X_{0}$ corresponds to the space of gauge
parameters or functions $f$, $X_{-1}$ is the space of gauge fields $A_a$ and $X_{-2}$ contains the left hand side of the
equations of motion of the gauge theory, $\mathpzc{F}(A)=0$. General elements in $X_{-2}$ are denoted by the letter $E$, i.e.,
\eq{      
\begin{matrix}   X_0\quad  &\quad X_{-1}\quad  &\quad  X_{-2} \\[0.2cm]
                        f    \quad  &\quad   A_a   \quad  &\quad  E_a
\end{matrix}\,.\label{fAE}
}
The gauge variations are defined in terms of the brackets $\ell_{n+1}(f,A^{n})\in X_{-1}$ as,
\eq{\label{var}
  \delta_{f}  A &=\sum_{n\ge 0}   {1\over n!}
      (-1)^{n(n-1)\over 2}\,
 \ell_{n+1}(f, \underbrace{ A, \dots, A}_{n \; {\rm times}} )\\&=\ell_1(f)+\ell_2(f,A)-\frac12\ell_3(f,A,A)+\dots\, .
 }
The equations of motion can be written as 
\eq{\label{calf}
{\mathpzc F}:=\underset{n \ge 1}{\sum} \;{1\over n!}
(-1)^\frac{n(n-1)}{2}\, \ell_n(A^n)=
   \ell_1(A)-{1\over 2} \ell_2(A^2)-{1\over 3!} \ell_3(A^3)+\ldots\; .
}

Using the L$_\infty$ relations (\ref{linftyrels}) one can show that the commutator of gauge variations yields \cite{HZ,Fulp:2002kk}
\eq{\label{closure}
\left[ \delta_{f}, \delta_{g} \right] A=
\delta_{-C(f,g,A) }\, A+ \delta^T_{-C(f,g,{\mathpzc F},A)}\,A
}
with
\eq{\label{clclosure}
C(f,g, A) & =\sum_{n\ge 0} {1\over n!}
            (-1)^{{n(n-1)\over 2}}
            \, \ell_{n+2}(f,g,  \underbrace{ A, \dots, A}_{n\;  {\rm times}})\,.
} 
and where the second term on the right hand side of \eqref{closure}
vanishes on-shell. It can be expanded as
\eq{
\label{defieomtrafo}
  C(f,g,{\cal F},A)=
  \ell_3(f,g,{\mathpzc  F})
+\ldots=
   \sum_{n\ge 0} {1\over n!}
            (-1)^{{n(n-1)\over 2}}
            \,\ell_{n+3}(f,g,{\mathpzc F},A^n)\,.
}
The gauge variation of the equation of motion ${\mathpzc F}$ reads
\eq{
\label{varF}
  \delta_{f}  {\mathpzc F} &=\ell_2(f,{\mathpzc F})+ \ell_3(f,{\mathpzc F},A)-{1\over 2} \ell_4(f,{\mathpzc F},A^2)+\ldots
}
reflecting that, as opposed to the gauge field $A$, it transforms covariantly.  

The same gauge algebra may correspond to the different gauge theories. So, following \cite{HZ} it is convenient to introduce two different L$_\infty$ algebras. The first of them denoted by L$^{gauge}_\infty$ describes only the gauge transformations of fields $A_a$ and is concentrated in two component space $X^{gauge}=X_{0}\oplus X_{-1} $. The second, L$^{full}_\infty$ is concentrated in three component space, $X^{full}=X_{0}\oplus X_{-1}\oplus X_{-2} $, and includes the information about the dynamics. This can be extended further as in \cite{Saemann}, adding the fourth component $X_{-3}$ containing the Noether identities, etc. For example, to define the L$^{gauge}_\infty$ algebra corresponding to the abelian gauge transformation, $\delta_f A_a=\partial_a f$, one sets $\ell_1(f)=\partial_a f$ and all other brackets vanishing. The L$^{full}_\infty$ algebra corresponding to the abelian Chern-Simons theory is defined by setting $\ell_1(A)^a=\varepsilon^{abc}\partial_bA_c$, while for Yang-Mills we define $\ell_1(A)^a=\Box A^a-\partial^a(\partial\cdot A)$. In both cases it is obvious that, $\ell_1(\ell_1(f))=0$. 

We stress that in principle, the L$_\infty$ algebra may have an infinite number of the brackets $\ell_n$, which however, are not arbitrary, since they should satisfy L$_\infty$ relations (\ref{linftyrels}). As it was already mentioned in the introduction the idea of the L$_\infty$ bootstrap approach consists in representing the original undeformed gauge theory together with a deformation as a part of a new L$_\infty$ structure by setting initial brackets and solving L$_\infty$ relations to determine the  L$_\infty$ algebra, which corresponds to the consistent deformation of the original theory.

\section{Non-commutative deformation of the abelian gauge transformations}

In this section we discuss a non-commutative deformation of the abelian L$^{gauge}_\infty$ algebra defined by the bracket, $\ell_1(f)=\partial_a f$. A deformation is introduced through the star commutator of functions which, from the consideration of anti-symmetry, should be assigned to the bracket,
\eq{\label{starcom}
\ell_2(f,g)=i[f,g]_\star\,.
}
For the Hermitean associative star product\footnote{These requirements to the star product are justified by the physical applications.} an expression for the star commutator up to the third order in $\Theta$ reads \cite{KV},
\begin{eqnarray}
&&i[f,g]_\star=-\Theta^{kl}\,\partial_kf\partial_l\,g  \notag \\
&&+\frac{1}{12}\,\left[\Theta ^{nl}\partial _{l}\Theta ^{mk}\partial _{n}\partial
_{m}\Theta ^{ij}\left( \partial _{i}f\partial _{j}\partial _{k}g-\partial
_{i}g\partial _{j}\partial _{k}f\right)\right.   \notag \\
&&+ \frac{1}{2}  \Theta^{nk}\partial_n\Theta^{jm}\partial_m\Theta^{il}
\left(\partial_i\partial_jf\partial_k\partial_lg-\partial_i\partial_jg\partial_k\partial_lf\right)   \notag \\
&&+\Theta ^{ln}\partial _{l}\Theta ^{jm}\Theta ^{ik}\left(
\partial _{i}\partial _{j}f\partial _{k}\partial _{n}\partial _{m}g-\partial
_{i}\partial _{j}g\partial _{k}\partial _{n}\partial _{m}f\right)
\notag \\
&&+\frac{1}{2}\Theta ^{jl}\Theta ^{im}\Theta ^{kn}\partial _{i}\partial
_{j}\partial _{k}f\partial _{l}\partial _{n}\partial _{m}g  \notag \\
&& +\left.\frac{1}{2}\Theta ^{nk}\Theta ^{ml}\partial _{n}\partial _{m}\Theta
^{ij}\left( \partial _{i}f\partial _{j}\partial _{k}\partial _{l}g-\partial
_{i}g\partial _{j}\partial _{k}\partial _{l}f\right)\right]+\mathcal{O}\left( \Theta ^{5}\right)~.   \label{star3}
\end{eqnarray}
In principle, the higher order terms in the star commutator can be constructed following the prescription of the Formality Theorem \cite{Kontsevich} and involves the Kontsevich graphs and the Kontsevich weights. The construction of the Kontsevich graphs is straightforward, however to the best of our knowledge there is no regular way of the computing the Kontsevich weights corresponding to the Kontsevich wheel diagrams. So, for arbitrary non-commutativity parameter $\Theta(x)$ an explicit all orders expression for the star commutator (\ref{star3}) is not known. 

In this work we will consider the limit of slowly varying, but not necessarily small fields, i.e., we discard the higher derivatives terms, like $\partial f \partial\partial g$, etc., in the star commutator and approximate it by the Poisson bracket, 
\eq{\label{ell2a}
\ell_2(f,g)\approx -\Theta^{ab}(x)\,\partial_a f\,\partial_b g=-\{f,g\}\,.
}
This is a ``self-consistent'' approximation of non-commutativity. If we work with the NC deformations induced by the associative star product, the star commutator (\ref{starcom}) satisfies the Jacobi identity, so does the corresponding Poisson bracket (\ref{ell2a}). Below we will see that in this case the brackets of the type $\ell_{n+2}(f,g,A^n)$, $n>0$, can be taken to be zero.

\subsection{Leading order contribution}

Having non-vanishing brackets $\ell_1(f)$ and $\ell_2(f,g)$, one has to check the L$_\infty$ relation, ${\cal J}_{2}(f,g)=0$, which reads,
\begin{eqnarray}\label{3.3}
  \ell_1( \ell_2 (f,g))&= -\{\overbrace{\partial_a  f}^{\in X_{-1}}, g\} -\{f, \overbrace{\partial_a g}^{\in X_{-1}}\}
    - (\partial_a \Theta^{ij})\, \partial_i f \partial_j g \\
       &= \ell_2(\ell_1(f),g)+\ell_2(f,\ell_1(g))\,,\nonumber
\end{eqnarray}
and involves the yet undetermined bracket $\ell_2(f,A)$. It means that now the identity, ${\cal J}_{2}(f,g)=0$, becomes an equation on $\ell_2(f,A)$. Solving this equation one may proceed to the next L$_\infty$ relation, ${\cal J}_{3}(f,g,h)=0$, and define the next bracket $\ell_3(f,g,A)$, etc.  

Let us see how it works on practice. 
From (\ref{3.3}) one finds
\begin{equation}\label{e.3.2}
\ell_2(f,A)=-\{f,A_a\} -{1\over 2} (\partial_a \Theta^{ij})\, \partial_i f A_j\,.
\end{equation}
Note that the solution is not unique, one may also set, e.g.,
\begin{equation}
\ell_2^\prime(f,A)= \ell_2(f,A)+s^{ij}_a(x)\, \partial_i f A_j\,, 
\end{equation}
with $s^{ij}_a(x)=s^{ji}_a(x)$. By the definition of L$_\infty$, $ \ell^\prime_2(A,f):=-\ell^\prime_2(f,A)$. The symmetry of $s^{ij}_a(x)$ implies that this choice of the bracket $\ell_2^\prime(f,A)$ also satisfies the equation (\ref{3.3}). However, the symmetric part $s^{ij}_a(x)\, \partial_i f \,A_j$ can always be ``gauged away'' by L$_\infty$-quasi-isomorphism, physically equivalent to a Seiberg-Witten map \cite{SW}, see  \cite{BBKT} for more details.

\subsection{Next to the leading order}

Then we have to define the bracket $\ell_3(f,g,A)$ from the identity ${\cal J}_{3}(f,g,h)=0$. Taking into account that according to (\ref{degree}) the bracket $\ell_3(f,g,h)$ belongs to the space $X_1$ which is empty in our construction, i.e., $\ell_3(f,g,h)=0$, one writes
\begin{eqnarray}
{\cal J}_{3}(f,g,h):=&&\ell_2(\ell_2(f,g),h)+\ell_2(\ell_2(g,h),f)+\ell_2(\ell_2(h,f),g)+\label{j30}\\
&&\ell_3(\ell_1(f),g,h)+\ell_3(f,\ell_1(g),h)+\ell_3(f,g,\ell_1(h))=0\ .\nonumber
\end{eqnarray}
The first line in the above expression is a Jacobiator for the bracket $\ell_2(f,g)$ defined in (\ref{ell2}), 
\begin{eqnarray}\label{ji}
&&\ell_2(\ell_2(f,g),h)+\ell_2(\ell_2(g,h),f)+\ell_2(\ell_2(h,f),g)=\\
&& \{\{f,g\},h\}+\{\{h,f\},g\}+\{\{g,h\},f\}\equiv0\,.\nonumber
\end{eqnarray}
So we may just set, $\ell_3(A,f,g)=0$.

The next step is the crucial for the whole construction. We have to analyze the relation ${\cal J}_{3}(f,g,A)=0$, given by
\begin{eqnarray}
0&=\ell_2(\ell_2(A,f),g)+\ell_2(\ell_2(f,g),A)+\ell_2(\ell_2(g,A),f)+\\
&\phantom{=i}\ell_1(\ell_3(A,f,g))-\ell_3(A,\ell_1(f),g)-\ell_3(A,f,\ell_1(g)).\nonumber
\end{eqnarray}
For simplicity, we replace it with ${\cal J}_{3}(g,h,\ell_1(f))=0$, written in the form
\begin{eqnarray}
&&\ell_3(\ell_1(f),\ell_1(g),h)-\ell_3(\ell_1(f),\ell_1(h),g)=G(f,g,h)\,,\label{G3}\\
&&G(f,g,h):=\ell_1(\ell_3(\ell_1(f),g,h))\nonumber\\
&&+\ell_2(\ell_2(\ell_1(f),g),h)+\ell_2(\ell_2(g,h),\ell_1(f))+\ell_2(\ell_2(h,\ell_1(f)),g)\,.\nonumber
\end{eqnarray}
We will follow the logic of \cite{KV} for the solution of the above algebraic equation.
By construction, the equation (\ref{G3}) is antisymmetric with respect to the permutation of $g$ and $h$. The graded symmetry of the $\ell_3$ bracket,  $\ell_3(\ell_1(f),\ell_1(g),h)= \ell_3(\ell_1(g),\ell_1(f),h)$, implies the identity on the l.h.s. of (\ref{G3}):
\begin{eqnarray*}
&&\ell_3(\ell_1(f),\ell_1(g),h)-\ell_3(\ell_1(f),\ell_1(h),g)+\\&&\ell_3(\ell_1(h),\ell_1(f),g)-\ell_3(\ell_1(h),\ell_1(g),f)+\\
&&\ell_3(\ell_1(g),\ell_1(h),f)-\ell_3(\ell_1(g),\ell_1(f),h)\equiv0\,.
\end{eqnarray*}
Which in turn requires the graded cyclicity of r.h.s. of the eq. (\ref{G3}),
\begin{equation}
G(f,g,h)+G(h,f,g)+G(g,h,f)=0\,.\label{ccG3}
\end{equation}
The latter is nothing but the consistency condition for the eq. (\ref{G3}).

It is remarkable that the consistency condition (\ref{ccG3}) follows from the previously satisfied L$_\infty$ relations, namely ${\cal J}_{2}(f,g)=0$, and ${\cal J}_{3}(f,g,h)=0$. Indeed,
taking the definition of $G(f,g,h)$, one writes
\begin{eqnarray*}
&&G(f,g,h)+G(h,f,g)+G(g,h,f)=\\
&&\ell_2(\ell_2(\ell_1(h),f),g)+\ell_2(\ell_2(f,g),\ell_1(h))+\ell_2(\ell_2(g,\ell_1(h)),f)+\\
&&\ell_2(\ell_2(\ell_1(g),h),f)+\ell_2(\ell_2(h,f),\ell_1(g))+\ell_2(\ell_2(f,\ell_1(g)),h)+\\
&&\ell_2(\ell_2(\ell_1(f),g),h)+\ell_2(\ell_2(g,h),\ell_1(f))+\ell_2(\ell_2(h,\ell_1(f)),g)+\\
&&\ell_1(\ell_3(\ell_1(f),g,h))+\ell_1(\ell_3(f,\ell_1(g),h))+\ell_1(\ell_3(f,g,\ell_1(h)))\,.
\end{eqnarray*}
Using ${\cal J}_{2}(f,g)=0$, we may push $\ell_1$ out of the brackets and rewrite it as
\begin{eqnarray*}
&&\ell_1\big[\ell_2(\ell_2(f,g),h)+\ell_2(\ell_2(g,h),f)+\ell_2(\ell_2(h,f),g)+\big.\\
&&\ell_3(\ell_1(f),g,h)+\ell_3(f,\ell_1(g),h)+\ell_3(f,g,\ell_1(h))\big]=\\
&&\ell_1\left[{\cal J}_{3}(f,g,h)\right]\equiv0\,.
\end{eqnarray*}
Which means that the consistency condition (\ref{ccG3}) holds true as a consequence of the previously satisfied L$_\infty$ relations. 

Taking into account (\ref{ccG3}) one may easily check that the following expression (symmetrization in $f$ and $g$ of the r.h.s. of the eq. (\ref{G3})):
\begin{equation}
\ell_3(\ell_1(f),\ell_1(g),h)=-{1\over 6}\Big(G(f,g,h)+G(g,f,h)\Big) \,,
\end{equation}
has the required graded symmetry and solves, ${\cal J}_{3}(g,h,\ell_1(f))=0$. For $\ell_1(f)=A,$ and $\ell_1(g)=B$, one gets,
\begin{eqnarray}\label{e.3.9}
\ell_3(A,B,f)_a=-{1\over 6}\Big(G_a^{kij}+G_a^{kji}\Big) A_i
     B_j \partial_k f\,,
\end{eqnarray}
with
\begin{eqnarray}\label{e.3.10}
G_a^{kij}=  -
  \Theta^{im} \partial_m \partial_a \Theta^{jk} -{1\over
    2} \partial_a\Theta^{jm} \partial_m\Theta^{ki}-{1\over
    2} \partial_a\Theta^{km} \partial_m\Theta^{ij}\,.
\end{eqnarray}
At this point we would like to stress three main observations. First, the consistency condition (graded cyclicity) (\ref{ccG3}) holds true as a consequence of the L$_\infty$ relations. Second,
 one needs higher brackets to compensate the violation of the standard Leibniz rule, which is standard in deformation quantization. And the last one is that the order of the bracket $\ell_{n+1}(F, A^n)$ in gauge fields $A^n$ corresponds to the order of this brackets in the non-commutativity parameter $\Theta^n$.

\subsection{Higher brackets and recurrence relations}

Once the brackets $\ell_3(f,g,A)$ and $\ell_3(f,A,B)$ are determined we may proceed to the next L$_\infty$ relations and find the brackets with four, five, etc., entries. First let us discuss the relations with three gauge parameters, ${\cal J}_{n+3}(f,g,h,A^n)=0$. The relation with four entries, ${\cal J}_{4}(f,g,h,A)=0$, can be represented schematically as follows,
\eq{\label{h1}
{\cal J}_4:=\ell_1\ell_4-\ell_4\ell_1+\ell_3\ell_2-\ell_2\ell_3=0\,.
}
Recall that by the consideration of the degree, $\ell_2(f,g)\in X_0$, and $\ell_2(f,A)\in X_{-1}$, and also, $\ell_3(f,g,h)=0$, and $\ell_4(f,g,h,A)=0$, since the corresponding brackets belong to the space $X_1$ which is empty in our construction. Taking into account now that the bracket, $\ell_3(f,g,A)=0$, due to (\ref{j30}), the identity (\ref{h1}) becomes,
\begin{eqnarray}\label{F4}
\ell_4(\ell_1(f),g,h,A)+\ell_4(f,\ell_1(g),h,A)+\ell_4(f,g,\ell_1(h),A)=0\,.
\end{eqnarray}
The latter can be satisfied setting, $\ell_4(f,g,A,B)=0$. The same arguments show that if the bracket $\ell_2(f,g)$ satisfies the Jacobi identity (\ref{ji}), then the L$_\infty$ relations with three gauge parameters, ${\cal J}_{n+3}(f,g,h,A^n)=0$, are satisfied for, $\ell_{n+2}(f,g,A^n)=0$, with $n\geq1$. On the other hand, this means that the closure condition for the gauge algebra becomes
\begin{equation}\label{an2}
 [\delta_{f},\delta_g] A
                      =\delta_{\{f,g\} }A \, .
\end{equation}

To proceed with the relations with two gauge parameters, ${\cal J}_{n+2}(g,h,A^n)=0$, we replace them by the equations, ${\cal J}_{n+2}(g,h,\ell_1(f)^n)=0$, which in turn can be represented in the form
\begin{eqnarray}
\ell_{n+2}(\ell_1(f)^{n},\ell_1(g),h)-\ell_{n+2}(\ell_1(f)^{n},\ell_1(h),g)= G(f_1,\dots,f_n,g,h)\,,\label{eqn}
\end{eqnarray}
where the right hand side, $G(f_1,\dots,f_n,g,h)$, is defined in terms of the previously defined brackets $\ell_{m+2}(\ell_1(f)^{m},\ell_1(g),h)$, with $m<n$. It is symmetric in the first $n$ arguments and antisymmetric in the last two by the construction. The graded symmetry of $\ell_{n+2}(\ell_1(f)^{n},\ell_1(g),h)$ implies the non-trivial consistency condition (since $G(f_1,\dots,f_n,g,h)$ is symmetric in first $n$ arguments, one needs to check the cyclicity relation with respect to the permutation of the last three slots),
\begin{eqnarray}
&&G(f_1,\dots,f_n,g,h)+G(f_1,\dots,f_{n-1},g,h,f_n)\label{ccGn}\\
&&+G(f_1,\dots,f_{n-1},h,f_n,g)=0\,,\nonumber
\end{eqnarray}
which follows from the previous L$_\infty$ relations and can be proved by induction. The solution of the equation (\ref{eqn}) can be constructed taking the symmetrization of the r.h.s. in the first $n+1$ arguments.

The order by order in $\Theta$ calculations indicates the ansatz for the brackets,
\eq{\label{elln2}
\ell_{n+2}\left(f,A^{n+1}\right)_a=(n+1)! (-1)^{\frac{n(n-1)}{2}}\,\Gamma_a^{ki_1\dots i_{n+1}}\,\partial_kf\,A_{i_1}\dots A_{i_{n+1}}\,,
}
yielding following expression for the gauge variation,
\eq{ \label{h2}
\delta_f A_a=\partial_a f+\{A_a, f\}+ \Gamma^k_a(A)\,\partial_kf\,,
}
where 
\eq{\label{h3}
 \Gamma^k_a(A)=\sum_{n= 1}^\infty \Gamma^{k(n)}_a\,,\qquad  \Gamma^{k(n)}_a=\Gamma^{ki_1\dots i_n}_a(x)\,A_{i_1}\dots A_{i_n}\,.
 }
 According to the slowly varying field approximation we take into account only the leading order contribution in derivatives $\partial f$ and $\partial A$. But we do not restrict the orders in $\Theta$. 
 The order of the term $ \Gamma^{k(n)}_a$ in the gauge  fields $A$ coincides with the order of this term in  deformation parameter $\Theta$. To determine $\Gamma^{k}_a(A)$ we use the closure condition (\ref{an2}). One finds after simplification that,
 \eq{\label{h4}&\delta_f\left(\delta_gA_a\right)-\delta_g\left(\delta_fA_a\right)-\delta_{\{f,g\}} A_a=\\&\left(\frac{\delta \Gamma^l_a}{\delta A_k}-\frac{\delta \Gamma^k_a}{\delta A_l}-
 \left(\delta^b_a+\Gamma^b_a\right)\,\partial_b\Theta^{kl}\right.\\&-\left.\partial_b\Gamma^l_a\,\Theta^{bk}+\partial_b\Gamma^k_a\,\Theta^{bl}- \frac{\delta\Gamma^k_a}{\delta A^b}\,\Gamma^l_b+ \frac{\delta\Gamma^l_a}{\delta A^b}\,\Gamma^k_b\right)\,\partial_kf\,\partial_lg\,,
 }
 where we use the notation, $\partial_b\Gamma^{l(n)}_a=(\partial_b\Gamma^{li_1\dots i_n}_a)\,A_{i_1}\dots A_{i_n}$. So, the closure condition (\ref{an2}) yields the following equation on $\Gamma^k_a(x,A)$,
 \eq{\label{h5}
 \frac{\delta \Gamma^l_a}{\delta A_k}-\frac{\delta \Gamma^k_a}{\delta A_l}=& \left(\delta^b_a+\Gamma^b_a\right)\,\partial_b\Theta^{kl}+\partial_b\Gamma^l_a\,\Theta^{bk}-\partial_b\Gamma^k_a\,\Theta^{bl}\\
 &+ \frac{\delta\Gamma^k_a}{\delta A^b}\,\Gamma^l_b- \frac{\delta\Gamma^l_a}{\delta A^b}\,\Gamma^k_b\,.
 }
 
In what follows we will construct the perturbative in $\Theta$ solution of the above equation. We substitute the decomposition (\ref{h3}) in the equation (\ref{h5}) and compare the same orders in $\Theta$ from the left and from the right. In the first order in $\Theta$ we obtain,
\eq{\label{h6}
\frac{\delta \Gamma^{l(1)}_a}{\delta A^k}-\frac{\delta \Gamma^{k(1)}_a}{\delta A^l}=G^{kl(1)}_a:=\partial_a\Theta^{kl}\,,
}
yielding a solution, $ \Gamma^{k(1)}_a=-\partial_a\Theta^{ kl}A_l/2$, which corresponds to (\ref{e.3.2}). The second order in $\Theta$ gives,
\eq{\label{h7}
\frac{\partial \Gamma^{l(2)}_a}{\delta A^k}-\frac{\partial \Gamma^{k(2)}_a}{\delta A^l}=G^{kl(2)}_a:=-\frac12\,G_a^{kli}\,A_i\,,
}
where $G_a^{kli}$ was defined in (\ref{e.3.10}). The consistency condition for the equation (\ref{h7}) reads,
\eq{\label{cch7}
\frac{\delta G^{kl(2)}_a}{\delta A^i}+\frac{\delta G^{li(2)}_a}{\delta A^k}+\frac{\delta G^{ik(2)}_a}{\delta A^l}=0\,.
}
It is equivalent to the requirement (\ref{ccG3}) and is satisfied as a consequence of the L$_\infty$ construction as it was discussed in the previous subsection. A solution of the equation (\ref{h7}), 
\eq{\label{sh7}\Gamma^{k(2)}_a=-\frac13\, G^{kl(2)}_a\,A_l=\frac16\, G_a^{kli}A_lA_i\,,}
represents exactly the contribution of the bracket $\ell_3(f,A,A)$ defined in (\ref{e.3.9}) to the gauge variation (\ref{var}).

In the higher orders in the deformation parameter $\Theta$ the eq. (\ref{h5}) results in,
\eq{\label{h8}
&\frac{\partial \Gamma^{l(n+1)}_a}{\delta A^k}-\frac{\partial \Gamma^{k(n+1)}_a}{\delta A^l}=G^{kl(n+1)}_a\,,\\
& G^{kl(n+1)}_a:=\frac12\,\Gamma^{b(n)}_a\,\partial_b\Theta^{kl}+\partial_b\Gamma^{l(n)}_a\,\Theta^{bk}-\frac12\,\partial_a\Theta^{kb}\,\Gamma^{l(n)}_b\\
&+\sum_{m= 1}^{n-1}\frac{\delta\Gamma^{k(m+1)}_a}{\delta A^b}\,\Gamma^{l(n-m)}_b- (k\leftrightarrow l)\,.
}
The consistency condition in this case,
\eq{\label{cch8}
\frac{\delta G^{kl(n+1)}_a}{\delta A^i}+\frac{\delta G^{li(n+1)}_a}{\delta A^k}+\frac{\delta G^{ik(n+1)}_a}{\delta A^l}=0\,,
}
is equivalent to the relations (\ref{ccGn}) which follow from the L$_\infty$ construction. A solution of the equation (\ref{h8}),
\eq{\label{sh8}
\Gamma^{k(n+1)}_a=-\frac{1}{n+2}\,G^{kl(n+1)}_aA_l\,,
}
defines the functions $\Gamma^{l(n+1)}_a$ in terms of the previously determined $\Gamma^{l(m)}_a$ with, $m\leq n$. 

The situation here is quite similar to the construction of the star product in the deformation quantization. For the arbitrary non-commutativity parameter $\Theta^{ab}(x)$ the best we can do is to provide the recurrence relations (\ref{sh8}) for the construction of the gauge variation (\ref{h2}). However for the specific choices of $\Theta$, like e.g., the linear one, it is possible to address the question of the convergence of the series (\ref{h3}) and provide all orders explicit formula for $\Gamma^k_a(A)$ in (\ref{h2}).

\section{Non-commutative field dynamics and L$_\infty$ structure}  

In this Section we discuss the consistent deformation of the field dynamics in the bootstrap approach, considering the example of the non-commutative deformation of the abelian Chern-Simons theory. In this case we write the initial brackets as
\eq{\label{CSI}
           \ell_1(f) = \partial_a f\,,\qquad
           \ell_1(A)=\varepsilon^{abc} \, \partial_b A_c \ ,\qquad \ell_2(f,g)=- \{f,g\}\,.
}
The brackets $\ell_{n+1}(f,A^n)$ and $\ell_{n+2}(f,g,A^n)$ were determined in the Section 3. The rest of the brackets $\ell_n(A^n)$, $\ell_{n+2}(f,E,A^{n})$, and $\ell_{n+3}(f,g,E,A^{n})$, can be found from the corresponding l$_\infty$ relations.

\subsection{Leading order contribution}

The first new non-trivial L$_\infty$ relation is
\eq{\label{g1}
  {\cal J}_2(f,A):=  \ell_1(\ell_2(f,A))-\ell_2(\ell_1(f),A)-\ell_2(f,\ell_1(A))=0 \,,
}
which we rewrite as 
\eq{\label{g2}
\ell_2(\ell_1(f),A)+\ell_2(f,\ell_1(A))= \ell_1(\ell_2(f,A))\,.
}
In the r.h.s. the bracket, $\ell_2(f,A)\in X_{-1}$, is given by (\ref{e.3.2}), while $\ell_1(A)$ is determined in (\ref{CSI}), so 
\eq{\label{g3}
\ell_1(\ell_2(f,A))=&- \varepsilon^{abc}\{\partial_b f,A_c\}- \{f,\varepsilon^{abc}\partial_b A_c\}- \varepsilon^{abc} \partial_b\Theta^{kl}\partial_k f\partial_l A_c\\&-\frac{1}{2} \varepsilon^{abc} \partial_c\Theta^{kl}\partial_k \partial_bf A_l- \frac{1}{2} \varepsilon^{abc} \partial_c\Theta^{kl}\partial_k f\partial_b A_l\,.
 }
The brackets $\ell_2(\ell_1(f),A)$ and $\ell_2(f,\ell_1(A))$ in the l.h.s. of (\ref{g2}) need to be determined. The bracket $\ell_2(f,E)$ should be antisymmetric with respect to the permutation of its arguments, so we identify
\eq{\label{g4}
\ell_2(f,\ell_1(A))=- \{f,\varepsilon^{abc}\partial_b A_c\}\,,\qquad\mbox{thus}\qquad\ell_2(f,E)=-\{f,E_a\}\,.
}
The rest of the eq. (\ref{g2}) can be written in the form
\eq{\label{g5}
\ell_2(\ell_1(f),A)=Q_1^{abck}\,\partial_k f\,\partial_bA_c+S_1^{abck}\,A_k\,\partial_b\partial_cf+T_1^{abckl}\,\partial_kA_b\,\partial_l\partial_cf\,,
}
where the coefficient functions $P_1^{aijk}$, $Q_1^{aijk}$ and $R_1^{aijkl}$ are given by
\eq{\label{g6}
Q_1^{abck}=&\,\varepsilon^{acm}\partial_m\Theta^{kb}-\frac12\,\varepsilon^{abm}\partial_m\Theta^{kc}\,,\\
S_1^{abck}=&-\frac14\left(\varepsilon^{abm}\partial_m\Theta^{ck}+\varepsilon^{acm}\partial_m\Theta^{bk}\right)\,,\\
T_1^{abckl}=&\,-\varepsilon^{abc}\,\Theta^{kl}\,.
}

The solution of the equation (\ref{g5}) will be constructed following the logic of the previous section. There is a non-trivial consistency condition coming from the graded symmetry of the bracket $\ell_2$, which is satisfied as a consequence of the previously solved L$_\infty$ relations. The relation $ {\cal J}_2(f,\ell_1(g))=0$, can be written as
\eq{\label{ce1}
\ell_2(\ell_1(f),\ell_1(g))=  \ell_1(\ell_2(f,\ell_1(g))) \,.
}
The graded symmetry of $\ell_2$ bracket, \eq{\label{g7}\ell_2(\ell_1(f),\ell_1(g))=\ell_2(\ell_1(g),\ell_1(f))\,,} implies the consistency condition on the right hand side of (\ref{ce1}),
\eq{\label{cc3}
  \ell_1(\ell_2(f,\ell_1(g)))- \ell_1(\ell_2(g,\ell_1(f)))=\,0\,.
  }
  The later  is automatically satisfied due to L$_\infty$ relation, ${\cal J}_2(f,g)=0$, since
  \eq{\label{ce2}
  \ell_1(\ell_2(f,\ell_1(g)))- \ell_1(\ell_2(g,\ell_1(f)))=&\\
  \ell_1\left[\ell_1(\ell_2(f,g))-{\cal J}_2(f,g)\right]\equiv&\,0 \,.
}
In the specific case of the deformation of Chern-Simons theory, i.e., eq. (\ref{g3}) the relation (\ref{cc3}) implies
\eq{\label{g8}
&Q_1^{abck}\,\partial_k f\,\partial_b\partial_c g+S_1^{abck}\,\partial_kg\,\partial_b\partial_cf+T_1^{abckl}\,\partial_k\partial_bg\,\partial_l\partial_cf=\\
&Q_1^{abck}\,\partial_k g\,\partial_b\partial_c f+S_1^{abck}\,\partial_kf\,\partial_b\partial_cg+T_1^{abckl}\,\partial_k\partial_bf\,\partial_l\partial_cg\,,
}
which in turn yields the following relations between the coefficients $Q_1^{aibck}$, $S_1^{abck}$ and $T_1^{abckl}$:
\eq{\label{g9}
S_1^{abck}=Q_1^{a(bc)k}\,,\qquad\mbox{and}\qquad T_1^{abckl}=T_1^{acblk}\,.
}
We stress that these relations can be checked explicitly using just (\ref{g6}), however they follow from the L$_\infty$ algebra constructed so far. Using (\ref{g9}) the origynal equation (\ref{g5}) becomes
\eq{\label{g10}
\ell_2(\ell_1(f),A)=Q_1^{abck}\,\big(\partial_k f\,\partial_bA_c+A_k\,\partial_b\partial_cf\big)+T_1^{abckl}\,\partial_kA_b\,\partial_l\partial_cf\,,
}
implying the solution
\eq{\label{g11}
\ell_2(B,A)=&\,\,Q_1^{abck}\,\big(B_k \,\partial_bA_c+A_k\,\partial_bB_c\big)+T_1^{abckl}\,\partial_kA_b\,\partial_lB_c\\
=&-\ \varepsilon^{abc}\{A_b,B_c\}+\varepsilon^{acm}\partial_m\Theta^{kb}\big(A_k\partial_bB_c +B_k  \partial_b A_c \big)\\
&-\frac12\,\varepsilon^{abm}\partial_m\Theta^{kc}\big(A_k\partial_bB_c +B_k  \partial_b A_c \big)\,,
}
which is in the perfect agreement with our previous result \cite{BBKL}. 

\subsection{Next to the leading order}

At  this order there appears higher brackets $\ell_3$. The
expressions for $\ell_3(A, f,g)$ and $\ell_3(A,B,f)$ were found in the previous
Section. Taking into account that now $X_{-2}$ is non trivial, one
may also have non-vanishing brackets $\ell_3(E,f,g)\in X_{-1}$,
$\ell_3(E,A,f)\in X_{-2}$ and $\ell_3(A,B,C)\in X_{-2}$.

Let us start with $\ell_3(E,f,g)$. Such a term contributes to the
closure condition, ${\cal J}_3(f,g,A)=0$, which is however satisfied 
without it. Therefore, we can set $\ell_3(E,f,g)=0$.
Next  we consider the L$_\infty$ relation, ${\cal J}_3(E,f,g)=0$, i.e.,
\eq{
&0=\ell_2(\ell_2(E,f),g)+\ell_2(\ell_2(g,E),f)+\ell_2(\ell_2(f,g),E)\\[0,1cm]
&\phantom{=i}+\ell_3(E,\ell_1(f),g)+\ell_3(E,f,\ell_1(g))\,.
}
Since, $\ell_2(E,f)=\{E_a,f\}$,  by (\ref{g4}), the first line in the above equation vanishes and one derives,
\eq{
\ell_3(E,A,f)=0\, .
}

Finally, to define $\ell_3(A,B,C)$, one has to solve, ${\cal J}(A,B,f)=0$,
written as,
\eq{\label{g12}
\ell_3\left(A,B,\ell_1(f)\right)=&\,r_3(A,B,f)\,,\\
r_3(A,B,f)=&-\ell_1(\ell_3(A,B,f))-\ell_3(\ell_1(A),B,f)+\ell_3(A,\ell_1(B),f)\\
&-\ell_2(\ell_2(A,B),f)-\ell_2(\ell_2(f,A),B)+\ell_2(\ell_2(B,f),A)\,.
}
By the construction the r.h.s., $r_3(A,B,f)$, is symmetric with respect to the permutation of $A$ and $B$. As before, the graded symmetry of the $\ell_3$ bracket,
\eq{
\ell_3\left(A,\ell_1(g), \ell_1(f)\right)=\ell_3\left(A,\ell_1(f), \ell_1(g)\right)\,,
}
implies a non-trivial \emph{consistency condition} on the r.h.s. of (\ref{g12}),

\eq{\label{g13}
r_3(A,\ell_1(g),f)=r_3(A,\ell_1(f),g)\,.
}
Using the previously satisfied L$_\infty$ relations, ${\cal J}_2(f,g)=0$, and, ${\cal J}_2(A,f)=0$, one may check that,
\eq{
r_3(A,\ell_1(g),f)-r_3(A,\ell_1(f),g)=\ell_1\left({\cal J}_3(g,f,A)\right)-{\cal J}_3(\ell_1(A),g,f)\equiv 0\,.
}
Again the consistency condition (\ref{g13}) is satisfied as a consequence of L$_\infty$ construction.

Now let us discuss the solution of the eq. (\ref{g12}) for the non-commutative deformation of CS theory. The calculation of the r.h.s. is quite involved, but straightforward. We represent it as
\eq{\label{g14}
\ell_3\left(A,B,\ell_1(f)\right)=
&Q_2^{aijkl}\left(A_i\,\partial_jB_k\,\partial_l f+B_i\,\partial_jA_k\,\partial_l f\right)+\\&S_2^{aijkl}\left(A_i\,B_l\,\partial_j\partial_kf+B_i\,A_l\,\partial_j\partial_kf\right)+\\
&T_2^{aijklm}\left(\partial_i f\,\partial_jA_k\,\partial_lB_m+\partial_i f\,\partial_jB_k\,\partial_lA_m\right)+\\
&U_2^{aijklm}\left(A_i\,\partial_jB_k\,\partial_l\partial_m f+B_i\,\partial_jA_k\,\partial_l\partial_m f\right)\,,
}
where
\eq{\label{g15a}
Q_2^{aijkl}=\varepsilon^{abj}&\left(\frac12\,\Theta^{lm}\partial_b\partial_m\Theta^{ki}+\frac16\Theta^{km}\partial_b\partial_m\Theta^{il}+\frac16\Theta^{im}\partial_b\partial_m\Theta^{kl}+\right.\\
&\,\,\,\,\left.\frac16\partial_b\Theta^{km}\partial_m\Theta^{il}-\frac13\partial_b\Theta^{im}\partial_m\Theta^{kl}\right)+\\
\varepsilon^{abk}&\left(\Theta^{lm}\partial_b\partial_m\Theta^{ij}-\frac12\Theta^{jm}\partial_m\partial_m\Theta^{il}+\right.\\
&\,\,\,\,\left.\partial_b\Theta^{im}\partial_m\Theta^{jl}-\frac12\partial_b\Theta^{jm}\partial_m\Theta^{il}\right)+\\
\varepsilon^{abc}&\left(\frac12\partial_b\Theta^{ij}\partial_c\Theta^{kl}-\frac12\partial_b\Theta^{ik}\partial_c\Theta^{jl}\right)\,,\\
S_2^{aijkl}=\varepsilon^{abl}&\left(\frac16\Theta^{im}\partial_b\partial_m\Theta^{jk}+\frac13\partial_b\Theta^{im}\partial_m\Theta^{jk}\right)\,, \\
T_2^{aijklm}=\frac12\,&\varepsilon^{abk}\Theta^{jl}\partial_b\Theta^{im}\,,\\
U_2^{aijklm}=\frac12\,&\varepsilon^{abk}\Theta^{jm}\partial_b\Theta^{il}-\frac12\,\varepsilon^{abm}\Theta^{jl}\partial_b\Theta^{ik}\,.
}
The equation (\ref{g13}) implies the following relations on the coefficient functions
\eq{\label{g17}
&Q_2^{aijkl}=Q_2^{aljki}\,,\\
&Q_2^{aijkl}=S_2^{aijkl}+S_2^{aljki}\,,\\
&U_2^{aijk(lm)}=T_2^{aijk(lm)}+T_2^{ai(lm)jk}\,.
}
We stress that the above relations are not manifest from the explicit form of the coefficient functions $P_2^{aijkl}$, $Q_2^{aijkl}$, $R_2^{aijklm}$ and $S_2^{aijk(lm)}$ given by (\ref{g15a}) correspondingly. They follow from the L$_\infty$ relations, ${\cal J}_3(g,f,A)=0$, ${\cal J}_3(E,g,f)=0$, etc., which were also used to obtain the eq. (\ref{g13}). The situation here is absolutely the same as in the previous Section for the construction of L$^{gauge}_\infty$-algebra. The solution of the L$_\infty$ relations in each given order $n$ imply the non-trivial consistency conditions, which in turn are satisfied due to the previously solved lower order L$_\infty$ relations.

The following expression
\eq{\label{g18}
\ell_3\left(A,B,C\right)=
\frac12Q_2^{aijkl}&\left(A_i\,\partial_jB_k\,C_l+C_i\,\partial_jA_k\,B_l+B_i\,\partial_jC_k\,A_l+\right.\\
&\,\,\left.C_i\,\partial_jB_k\,A_l+B_i\,\partial_jA_k\,C_l+A_i\,\partial_jC_k\,B_l\right)+\\
T_2^{aijklm}&\left(A_i\,\partial_jB_k\,\partial_lC_m+C_i\,\partial_jA_k\,\partial_lB_m+B_i\,\partial_jC_k\,\partial_lA_m+\right.\\
&\,\,\left.C_i\,\partial_jB_k\,\partial_lA_m+B_i\,\partial_jC_k\,\partial_lB_m+A_i\,\partial_jC_k\,\partial_lB_m\right)\,,
}
by construction is symmetric in all arguments and due to the relations (\ref{g17}) satisfies the equation (\ref{g14}). 

\subsection{Higher order relations}

Following the same logic as in the beginning of the Section 3.3 we conclude that the L$_\infty$ relations of the type, ${\cal J}_n(A^{n-3}E,f,g)=0$, are satisfied for, $\ell_n(A^{n-2},E,f)=0$, for $n>0$. Thus, we conclude that the condition (\ref{varF}) on the gauge variation of the field equation, ${\mathpzc F}=0$, defined by (\ref{calf}) becomes
\eq{
\label{e1}
  \delta_{f}  {\mathpzc F} &=\ell_2(f,{\mathpzc F})= \{{\mathpzc F},f\}\,.
}
That is, the gauge variation of the field equation is proportional to the field equation itself, i.e., it is gauge invariant on-shell. The proper field equations, ${\mathpzc F}=0$, are constructed from the brackets $\ell_{n}(A^n)$ which should be determined from the L$_\infty$ relations, ${\cal J}_n(f,A^{n-1})=0$. The latter can be schematically represented in the form
\eq{\label{gn}
\ell_n\left(A^{n-1}, \ell_1(f)\right)=r_n\left(A^{n-1},f\right)\,,
}
where the r.h.s. $r_n\left(A^{n-1},f\right)$ written in terms of the lower order brackets $\ell_m$, $m<n$, by the construction is symmetric in first $n-1$ arguments. Like in the case of first two orders given by the equations (\ref{g5}) and (\ref{g14}) correspondingly, the eq. (\ref{gn}) has a non-trivial consistency condition. The graded symmetry of the $\ell_n$-bracket on the l.h.s. of (\ref{gn}) implies the relation,
\eq{\label{cc1}
r_n\left(A^{n-2},\ell_1(g),f\right)=r_n\left(A^{n-2},\ell_1(f),g\right)\,.
}
This relation can be proved by induction. To construct a solution of the equation (\ref{gn}) one may follow the prescription of the previous subsection, in particular (\ref{g14}) and (\ref{g18}).

The form of the lower order brackets indicates the anzats,
\eq{\label{ellnA}
\ell_n\left(A^n\right)^a=n!(-1)^{\frac{n(n-1)}{2}}\,\left( P^{abc(n-1)}(A)\,\partial_bA_c+R^{abc(n-2)}\left(A\right)\,\left\{A_b,A_c\right\}\right)\,,
}
where,
\eq{
&P^{abc(0)} =\varepsilon^{abc}\,,\qquad P^{abc(n)}= P^{abci_1\dots i_n}(x)\,A_{i_1}\dots A_{i_n}\,,\\
& R^{abc(0)}=\frac12\,\varepsilon^{abc}\,,\qquad  R^{abc(n)}=R^{abci_1\dots i_n}(x)\,A_{i_1}\dots A_{i_n}\,.
}
The latter in turn means that for the l.h.s. of the field equation one writes,
\eq{
\label{e2}
   {\mathcal  F}^a=P^{abc}\left(A\right)\,\partial_b A_c+R^{abc}\left(A\right)\,\left\{A_b,A_c\right\}\,,
}
with
\eq{\label{ic1}
P^{abc}\left(A\right)=\sum_{n= 0}^\infty P^{abc(n)}\,,\qquad \mbox{and}\qquad R^{abc}\left(A\right)=\sum_{n= 0}^\infty R^{abc(n)}\,,
}
The notations used here are similar to those in (\ref{h3}). The order of the term $P^{abc(n)}$ in the gauge fields $A^n$ coincides with the order of this term in the deformation parameter $\Theta^n$. By construction, $R^{abc}\left(A\right)$ should be antisymmetric in $b$ and $c$, since it is contracted with the Poisson bracket, $\left\{A_b,A_c\right\}$.

To determine the functions $P^{abc}(A)$ and $R^{abc}(A)$ we use the gauge covariance condition (\ref{e1}).  Introducing the notation
\eq{\label{delf}\delta_{f}=\bar\delta_{f}+\{\,\cdot\,,f\}\,,}
 one obtains in the l.h.s. of (\ref{e1}):
\eq{
\label{e5}
  \delta_{f}  {\mathcal F}_a=&\big(\bar\delta_{f}\,P^{abc}\big)\,\partial_bA_c+P^{abc}\,\partial_b\big(\bar\delta_{f}\,A_c\big)+P^{abc}\,\partial_b\Theta^{kl}\,\partial_kA_c\,\partial_l f\\
  &+P^{abc}\,\{A_c,\partial_bf\}+\{P^{abc}\,\partial_bA_c,f\}\\
  &+\big(\bar\delta_{f}\,R^{abc}\big)\,\left\{A_b,A_c\right\}+\{R^{abc},f\}\,\left\{A_b,A_c\right\}\\
  &+2\,R^{abc}\,\left\{\bar\delta_{f}\,A_b,A_c\right\}+R^{abc}\,\left\{\{A_b,f\},A_c\right\}+R^{abc}\,\left\{A_b,\{A_c,f\}\right\}\,.
  }
While the r.h.s. of (\ref{e1}) is just given by
\eq{
\label{e6}
\{P^{abc}\,\partial_b A_c+R^{abc}\,\left\{A_b,A_c\right\},f\}\,.
}
Taking into account that due to Jacobi identity,
\eq{
R^{abc}\,\big(\{A_b,\{A_c,f\}\}+\{A_c,\{f,A_b\}\}+\{f,\{A_b,A_c\}\}\big)\equiv0\,,
}
the eq. (\ref{e1}) for the ansatz (\ref{e2}) becomes
\eq{
\label{e7}
&\big(\bar\delta_{f}\,P^{abc}\big)\,\partial_bA_c+P^{abc}\,\partial_b\big(\bar\delta_{f}\,A_c\big)+P^{abc}\,\partial_b\Theta^{kl}\,\partial_kA_c\,\partial_l f\\
  &+P^{abc}\,\{A_c,\partial_bf\}+\big(\bar\delta_{f}\,R^{abc}\big)\,\left\{A_b,A_c\right\}+2\,R^{abc}\,\left\{\bar\delta_{f}\,A_b,A_c\right\}=0\,.
  }
We set separately
\eq{
\label{e8}
\big(\bar\delta_{f}\,P^{abc}\big)\,\partial_bA_c+P^{abc}\,\partial_b\big(\bar\delta_{f}\,A_c\big)+P^{abc}\,\partial_b\Theta^{kl}\,\partial_kA_c\,\partial_l f=0\,,
}
and
\eq{
\label{e9}
P^{abc}\,\{A_c,\partial_bf\}+\big(\bar\delta_{f}\,R^{abc}\big)\,\left\{A_b,A_c\right\}+2\,R^{abc}\,\left\{\bar\delta_{f}\,A_b,A_c\right\}=0\,.
  }
  
Let us analyze first the equation (\ref{e8}). We remind that by (\ref{h2}) and (\ref{delf}), $\bar\delta_{f}A_a=\partial_a f+\Gamma_a^k(A)\,\partial_kf$. So the equation (\ref{e8}) can be written as
\eq{\label{e8a}
&\left[\frac{\delta P^{abc}}{\delta A^l}\,\left(\delta^k_l+\Gamma^k_l\right)+P^{abl}\,\frac{\delta\Gamma^k_l}{\delta A^c}+P^{alc}\,\partial_l\Theta^{bk}\right]\,\partial_bA_c\,\partial_kf+\\
&P^{abc}\,\left(\delta^k_c+\Gamma^k_c\right)\,\partial_b\partial_kf=0\,.
}
Thus, we obtain two separate conditions on $P^{abc}$:
\eq{\label{e8a1}
\frac{\delta P^{abc}}{\delta A^l}\,\left(\delta^k_l+\Gamma^k_l\right)+P^{abl}\,\frac{\delta\Gamma^k_l}{\delta A^c}+P^{alc}\,\partial_l\Theta^{bk}=0\,,
}
and
\eq{\label{e8a2}
P^{abc}\,\left(\delta^k_c+\Gamma^k_c\right)+P^{akc}\,\left(\delta^b_c+\Gamma^b_c\right)=0\,.
}
From this point we act in the same way as in the Section 3.3, we will construct the perturbative in $\Theta$ solution of the above equations using the expression for $\Gamma^k_l$ obtained in the previous Section. The equation (\ref{e8a1}) in the first order in $\Theta$ reads,
\eq{\label{P1}
-\frac{\delta P^{abc(1)}}{\delta A^k}=Q^{abck(1)}:=\varepsilon^{acm}\partial_m\Theta^{kb}-\frac12\,\varepsilon^{abm}\partial_m\Theta^{kc}\,.
}
The r.h.s. of the above equation is exactly the coefficient $Q_1^{abck}$ determined in (\ref{g6}).
Its solution, 
\eq{\label{sP1}
P^{abc(1)}=-Q^{abck(1)}A_k\,,
}
reproduces the contribution to the equations of motion (\ref{calf}) from the corresponding part of the bracket $\ell_2(A,A)$ defined in (\ref{g11}). This choice for $P^{abc(1)}$ also satisfies the equation (\ref{e8a1}) up to the first order in $\Theta$. It can be checked explicitly, but also follows from the first of the equations (\ref{g9}), that is from the L$_\infty$ construction. 

In the higher orders in $\Theta$ the equation (\ref{e8a1}) results in,
\eq{\label{e10}
-\frac{\delta P^{abc(n+1)}}{\delta A^k}=&\,\,Q^{abck(n+1)}\,,\\
Q^{abck(n+1)}:=&\sum_{m= 1}^n \frac{\delta P^{abc(n+1-m)}}{\delta A^l}\,\Gamma^{k(m)}_l\\+&\sum_{m= 0}^n P^{abl(n-m)}\,\frac{\delta \Gamma^{k(m+1)}}{\delta A^c}+P^{alc(n)}\partial_l\Theta^bk\,.
} 
The consistency condition for this equation reads,
\eq{\label{cce10}
\frac{\delta Q^{abck(n+1)}}{\delta A^l}=\frac{\delta Q^{abcl(n+1)}}{\delta A^k}\,.
}
It follows from the relation (\ref{cc1}). A solution is
\eq{\label{se10}
P^{abc(n+1)}=-\frac{1}{n+1}\,Q^{abck(n+1)}A_k\,.
}
As before, for arbitrary non-commutativity parameter $\Theta^{kl}(x)$ we can only provide the recurrence relations (\ref{se10}) for definition of the functions $P^{abc(n+1)}$. In the next Section we will construct an explicit all orders expression for $P^{abc}(A)$ for linear $\Theta$.

Once $P^{abc}(A)$ is found, from (\ref{e8a2}) one determines,
\eq{\label{s9}
R^{abc}=\frac12\,\left[\left(\delta+\Gamma\right)^{-1}\right]^b_k\,P^{akc}\,.
}
The antisymmetry, $R^{abc}=-R^{acb}$, follows from (\ref{e8a2}).

\section{Lie-algebra like deformation}

The main goal of this Section is to work out the most simple and at the same time non-trivial situation taking the non-commutativity parameter $\Theta$ to be linear function of the coordinates and satisfying the Jacobi identity. Physically it corresponds, for example, to the $Q$-flux backgrounds in open string theory \cite{LMPS}. 

\subsection{NC $su(2)$-like deformation}

We choose the non-commutativity parameter, $\Theta^{ij}(x)=2\,\theta\,\varepsilon^{ijk} x_k$, which correspond to the rotationally invariant $3$d NC space \cite{Hammou:2001cc,GraciaBondia:2001ct,Vitale:2012dz,Galikova:2013zca,Kupriyanov:2012nb}. For the brevity of the calculations in most cases we will suppress the symbol $\theta$ in this and the following subsections. However, we will restore $\theta$ in the Subsection 5.3 where we provide the summary of the main findings of this Section. The corresponding Poisson bracket is
\begin{equation}\label{su2}
\{f,g\}=2\,\varepsilon^{ijk} x_k\ \partial_if\ \partial_jg\,.
\end{equation}
For the first two brackets with one gauge parameter one finds, 
\eq{\label{var1}
  \ell_2(f,A)_a &=\{A_a,f\}+\varepsilon_a{}^{bc}A_b\partial_cf\\\ell_3(f,A,A)_a&=-\frac23\left(\partial_af A^2-\partial_bfA^bA_a\right)\, ,
 }
 with  $A^2=A_bA^b$.
Then, using the recurrence relations (\ref{sh8}) we observe that the brackets $\ell_{n+3}(f,A^n)$ with the odd $n$ vanish, while for even $n$ they have the structure
\eq{\label{var2}
  \ell_{n+3}(f,A^n)=\left(\partial_af A^2-\partial_bfA^bA_a\right)\chi_n(A^2)\,,
 }
 for some monomial function $\chi_n(A^2)$. The combination of (\ref{var1}) and (\ref{var2}) in (\ref{h2}) results in the following ansatz for the gauge variation:
\begin{equation}
 \delta_{f}  A_a=\partial_af+\{A_a,f\}+\varepsilon^{abc}A_b\partial_cf+\left(\partial_af A^2-\partial_bfA^bA_a\right)\chi(A^2)\,,\label{an1}
\end{equation}
where the function, $\chi(A^2)=\sum_n \chi_n(A^2),$ should be determined from the closure condition (\ref{an2}).

For the convenience of the reader here we repeat the calculation (\ref{h4}) for the specific form of the gauge variation (\ref{an1}),
\begin{eqnarray}
&&\delta_f\left(\delta_gA_a\right)-\delta_g\left(\delta_fA_a\right)-\delta_{\{f,g\}} A_a=\label{n2}\\
&&\{\delta_fA_a,g\}+\varepsilon^{abc}\delta_fA_b\partial_cg+\left(2\partial_agA_b\delta_fA^b-\partial_bg\delta_fA^ba_a-\partial_bgA^b\delta_fA_a\right)\chi(A^2)\nonumber\\
&&+\left(\partial_ag A^2-\partial_bgA^bA_a\right)\chi^\prime(A^2)2A_c\delta_fA^c\nonumber\\
&&-\{\delta_gA_a,f\}-\varepsilon^{abc}\delta_gA_b\partial_cf-\left(2\partial_afA_b\delta_gA^b-\partial_bf\delta_gA^ba_a-\partial_bfA^b\delta_gA_a\right)\chi(A^2)\nonumber\\
&&-\left(\partial_af A^2-\partial_bfA^bA_a\right)\chi^\prime(A^2)2A_c\delta_gA^c\nonumber\\
&&-\partial_a\{f,g\}-\{A_a,\{f,g\}\}-\varepsilon^{abc}A_b\partial_c\{f,g\}-\left(\partial_a\{f,g\} A^2-\partial_b\{f,g\} A^bA_a\right)\chi(A^2)\,.\nonumber
\end{eqnarray}
After tedious but straightforward calculations we rewrite the r.h.s. of (\ref{n2}) as
\begin{equation}
\left[\partial_ag\partial_bfA^b-\partial_af\partial_bgA^b\right]\left(1+3\chi(A^2)+A^2\chi^2(A^2)+2A^2\chi^\prime(A^2)\right)\,.
\end{equation}
That is, requiring that
\begin{equation}\label{ode}
2t\chi^\prime(t)+1+3\chi(t)+t\chi^2(t)=0\,,\qquad \chi(0)=-\frac{1}{3}\,,
\end{equation}
we will obtain zero in the r.h.s. of (\ref{n2}). 
The solution of (\ref{ode}) is
\begin{equation}\label{solode}
\chi(t)=\frac1t\,\left(\sqrt{t}\cot\sqrt{t}-1\right)\,.
\end{equation}

Thus, we have obtained in (\ref{an1}) and (\ref{solode}) with, $t=\theta^2A^2$, an explicit form of the non-commutative $su(2)$-like deformation of the abelian gauge transformations in the slowly varying field approximation. Following the lines described in \cite{Kupriyanov:2015uxa} this result can be generalized for the non-commutative deformations along any linear Poisson structure $\Theta^{ij}(x)$.

\subsection{Non-commutative Chern-Simons theory}

The initial data in this case were specified in (\ref{CSI}). The brackets $\ell_2(A,A)$ and $\ell_3(A^3)$ were calculated in (\ref{g11}) and (\ref{g18}) correspondingly. The resulting expression for the NC $su(2)$-like deformation of the abelian Chern-Simons equations of motion up to the order ${\cal O}(\Theta^3)$ is given by:
\eq{\label{nceom}
\mathpzc F^a:=&\,\varepsilon^{abc}\partial_bA_c+\frac12\,\varepsilon^{abc}\{A_b,A_c\}+\theta(2A^b\partial^aA_b- A^a\partial^bA_b-A^b\partial_bA^a)\\
&+\theta^2\left[-\frac83\,\varepsilon^{abc}A^2\partial_bA_c+\frac{2}{3}\,\varepsilon^{abm}A_m\,A^c\,\partial_bA_c-2\,\varepsilon^{acm}A_mA^b\partial_bA_c\right.\\
&+\left.2\,\varepsilon^{bcm}A_m\,A^a\partial_bA_c\right]-\theta\{A^2,A_a\}+{\cal O}(\theta^3)=0\,.
}

\subsubsection*{Definition of the $P$-term}
  
Taking into account an explicit form of the gauge variation $\delta_f A_a$ given by (\ref{an1}), the equations (\ref{e8a1}) and (\ref{e8a2}) in case of linear $\Theta$ become,
\eq{
\label{e11}
&\frac{\delta P^{abc}}{\delta A_e}\,\left(1+A^2\,\chi\right)+\frac{\delta P^{abc}}{\delta A^m}\,\left(\varepsilon^{mne}A_n- A^m\,A^e\,\chi\right)\\
&+P^{abm}\,\varepsilon^{cem}+2\,P^{amc}\,\varepsilon^{bem} +2\,P^{abe}\,A^c\,\left(\chi+A^2\,\chi^\prime\right)\\
&-P^{abc}\,A^e\,\chi-P^{abm}\,A_m\,\delta^{ce}\,\chi-2\,P^{abm}\,A_m\,A^c\,A^e\,\chi^\prime=0\,,
}
and
\eq{
\label{e12}
&\big(P^{abc}+P^{acb}\big)\,\left(1+A^2\,\chi\right)-P^{abm}\,\varepsilon^{cnm}A_n-P^{acm}\,\varepsilon^{bnm}A_n\\
&-P^{abn}A_n\,A^c\,\chi-P^{acn}A_n\,A^b\,\chi=0\,.
}
Again the lower order brackets $\ell_n(A^n)$ indicate the anzatz
\eq{
\label{e3}
P^{abc}\left(A\right)=&\varepsilon^{abc}\,F\left(A^2\right)+\varepsilon^{abm}A_m\,A^c\,G\left(A^2\right)+\varepsilon^{acm}A_m\,A^b\,H\left(A^2\right)\\&+\varepsilon^{bcm}A_m\,A^a\,J\left(A^2\right)
+A^a\,A^b\,A^c\,K\left(A^2\right)\\&+A^a\,\delta^{bc}\,L\left(A^2\right)+A^b\,\delta^{ac}\,M\left(A^2\right)+A^c\,\delta^{ab}\,N\left(A^2\right)\,.  
}
The equation (\ref{e12}) implies the following relations on the coefficient functions:
\eq{\label{e12a}
& G+H\,\left(1+A^2\,\chi\right)-\chi\,F-M=0\,,\\
&K-\chi\,(L+M)-J-H=0\,,\\
&L\,\left(1+A^2\,\chi\right)+F+A^2\,J=0\,,\\
&M\,\left(1+A^2\,\chi\right)+N-F+A^2\,H=0\,.
}

Our strategy is to substitute (\ref{e3}) in (\ref{e11}) and collect the coefficients at the different powers of fields $A$, modulo the $A^2$. Starting with a quartic in $A$ contribution, $A^a\,A^b\,A^c\,A^e$, then cubic in $A$ structures, like $\varepsilon^{abm}A_m\,A^c\,A^e$, etc. down to the zero order in $A$ terms like $\delta^{ab}\,\delta^{ce}$. Equating to zero these coefficients we will obtain the system of differential equations on the coefficient functions $F,\dots,N$. We stress that there are algebraic relations involving the Levi-Civita tensors $\varepsilon^{abc}$ and vector fields $A^e$ described in the appendix. Using them we will reduce the number of different structures and thus the number of the equations on $F$, $G$, etc. These relations guarantee that the resulting system of differential equations is not overfull. The equation (\ref{e11}) does have the solution.

We start writing a term quartic in $A$ term in the l.h.s. of (\ref{e11}):
\eq{\label{e25}
A^a\,A^b\,A^c\,A^e\,\left[2\,K^\prime-2\,\chi\,K-2\,\chi^\prime\,\left(L+M+A^2\,K\right)+2\,A^2\,\chi^\prime\,K\right]\,.
}
The cubic in the field $A$ contribution is given by
\eq{\label{e26}
&\varepsilon^{abm}A_m\,A^c\,A^e\left[2\,G^\prime-\chi\,G+2\,A^2\,\chi^\prime\,G\right]+\\
&\varepsilon^{acm}A_m\,A^b\,A^e\left[2\,H^\prime-3\chi\,H\right]+\varepsilon^{bcm}A_m\,A^a\,A^e\left[2\,J^\prime-3\,\chi\,J\right]+\\
&\varepsilon^{aem}A_m\,A^b\,A^c\left[2\,\left(\chi+A^2\,\chi^\prime\right)\,H-K\right]+\\
&\varepsilon^{bem}A_m\,A^a\,A^c\left[2\,\left(\chi+A^2\,\chi^\prime\right)\,J+K\right]\,.
}
At this point for the first time we make use of the algebraic relation from the appendix to reduce the number of structures. Namely employing the identity
\eq{
\varepsilon^{aem}A_m\,A^b-\varepsilon^{bem}A_m\,A^a=-\varepsilon^{abe}A^2+\varepsilon^{abm}A_m\,A^e\,,
}
and setting, $J=-H$, one rewrites (\ref{e26}) as
\eq{\label{e26a}
&\varepsilon^{abm}A_m\,A^c\,A^e\left[2\,G^\prime-\chi\,G+2\,A^2\,\chi^\prime\,G+2\,\left(\chi+A^2\,\chi^\prime\right)\,H-K\right]+\\
&\varepsilon^{acm}A_m\,A^b\,A^e\left[2\,H^\prime-3\chi\,H\right]+\varepsilon^{bcm}A_m\,A^a\,A^e\left[2\,J^\prime-3\,\chi\,J\right]+\\
&\varepsilon^{abe}\,A^c\left[A^2\,K-2\,A^2\,\left(\chi+A^2\,\chi^\prime\right)\,H\right]\,.
}
We stress that now there has appeared the linear in $A$ contribution coming from the cubic ones.

We continue with the  quadratic in the fields $A$ terms in the l.h.s. of (\ref{e11}),
\eq{\label{e19}
&\delta^{ae}\,A^b\,A^c\left[(1+A^2\chi)\,K+G+2(\chi+A^2\chi^\prime)\,M\right]+\\
&\delta^{be}\,A^a\,A^c\left[(1+A^2\chi)\,K+G+2(\chi+A^2\chi^\prime)\,L\right]+\\
&\delta^{ce}\,A^a\,A^b\left[K-J-H-\chi(L+M)\right]+\\
&\delta^{bc}\,A^a\,A^e\left[2\,L^\prime-2\chi L+J\right]+\delta^{ac}\,A^b\,A^e\left[2\,M^\prime-2\chi M-H\right]+\\
&\delta^{ab}\,A^c\,A^e\left[2\,N^\prime-2G\right]+\varepsilon^{acm}A_m\varepsilon^{ben}A_n\,H-\varepsilon^{aem}A_m\varepsilon^{bcn}A_n\,J\,.
}
Using the identity (\ref{E5}) from the appendix which we write here for the convenience of the reader,
\eq{
\varepsilon^{acm}A_m\varepsilon^{ben}A_n&=\left(\delta^{ab}\,\delta^{ce}-\delta^{ae}\,\delta^{bc}\right)\,A^2\\
&+\delta^{bc}\,A^a\,A^e-\delta^{ce}\,A^a\,A^b-\delta^{ab}\,A^c\,A^e+\delta^{ae}\,A^b\,A^c\,,
} 
we rewrite (\ref{e19}) as
\eq{\label{e20}
&\delta^{ab}\,\delta^{ce}\,A^2\,\left[H-J\right]-\delta^{ae}\,\delta^{bc}\,A^2\,H+\delta^{ac}\,\delta^{be}\,A^2\,J+\\
&\delta^{ae}\,A^b\,A^c\left[(1+A^2\chi)\,K+G+2(\chi+A^2\chi^\prime)\,M+H\right]+\\
&\delta^{be}\,A^a\,A^c\left[(1+A^2\chi)\,K+G+2(\chi+A^2\chi^\prime)\,L-J\right]+\\
&\delta^{ce}\,A^a\,A^b\left[K-J-H-\chi(L+M)-H+J\right]+\\
&\delta^{bc}\,A^a\,A^e\left[2\,L^\prime-2\chi L+J+H\right]+\delta^{ac}\,A^b\,A^e\left[2\,M^\prime-2\chi M-H-J\right]+\\
&\delta^{ab}\,A^c\,A^e\left[2\,N^\prime-2G-H+J\right]\,.
}

At this point it is convenient to invert the order. First we will analize the zero order in $A$ contributions in the equation (\ref{e11}) and only then the linear in the fields $A$ terms. Taking into account the first line of (\ref{e20}) the zero order in $A$ terms in the l.h.s. of (\ref{e11}) are given by
\eq{\label{e13}
&\delta^{ac}\,\delta^{be}\,\left[ F+\left(1+A^2\,\chi\right)\,M+J\right]+\delta^{ae}\,\delta^{bc}\,\left[F+\left(1+A^2\,\chi\right)\,L-H\right]\\
&+\delta^{ab}\,\delta^{ce}\,\left[N-2F+H-J\right]\,.
}
The significant simplification occurs if we set
\eq{\label{HJ}
H=-J=0\,.
}
In order to the equation (\ref{e11}) be satisfied the coefficients at the different structures in the l.h.s. should be equal to zero. Thus from (\ref{e13}) we get
\eq{\label{e14}
L=M=-\frac{F}{1+A^2\,\chi}\,,\qquad\mbox{and}\qquad N=2\,F\,.
}
Equating to zero the coefficient at $\delta^{ce}\,A^a\,A^b$ in (\ref{e20}) one finds,
\eq{\label{e21a}
K=\chi\,(L+M)\,.
}

Now let us return to the linear in $A$ contributions to the left hand side of the equation (\ref{e11}). Taking into account (\ref{e26a}) it can be written as
\eq{\label{e15}
&\varepsilon^{abc}\,A^e\,\left[2\,F^\prime-\chi\,F\right]+\\
&\varepsilon^{abe}\,A^c\,\left[\left(1+A^2\,\chi\right)\,G+2\,N+2\,(\chi+A^2\,\chi^\prime)\,F\right.\\
&+\left.A^2\,K-2\,A^2\,\left(\chi+A^2\,\chi^\prime\right)\,H\right]+\\
&\varepsilon^{ace}\,A^b\,\left[\left(1+A^2\,\chi\right)\,H+M\right]+\varepsilon^{bce}\,A^a\,\left[\left(1+A^2\,\chi\right)\,J-L\right]+\\
&\varepsilon^{abm}\,A_m\,\delta^{ce}\,\left[G-\chi\,F\right]+\varepsilon^{acm}\,A_m\,\left(1+A^2\,\chi\right)\,H-\varepsilon^{aem}\,A_m\,\delta^{bc}\,L+\\
&\varepsilon^{bcm}\,A_m\,\delta^{ae}\,\left(1+A^2\,\chi\right)\,J+\varepsilon^{bem}\,A_m\,\delta^{ac}\,M\,.
}
Here we recall that because of the algebraic identities from the appendix not all structures in the above expression are independent. Now using these identities and previously defined  coefficients we will reduce the number of terms in (\ref{e15}). First, using (\ref{E4}) and (\ref{e14}) we get rid of the terms,
\begin{equation*}
-\varepsilon^{aem}\,A_m\,\delta^{bc}\,L+\varepsilon^{bem}\,A_m\,\delta^{ac}\,M\,,
\end{equation*}
substituting them with,
\begin{equation*}
\varepsilon^{abe}\,A^c\,L-\varepsilon^{abm}A_m\,\delta^{ec}\,L\,.
\end{equation*}
Then we utilize the identity (\ref{E1}) to convey the terms,
$$
\varepsilon^{ace}\,A^b\,M-\varepsilon^{bce}\,A^a\,L\,,
$$
through the $$-\varepsilon^{abc}\,A^e\,L+\varepsilon^{abe}\,A^c\,L\,.$$
We use that, $H=-J=0$, from (\ref{HJ}), and also notice that due to (\ref{e14}) and (\ref{e21a}) the coefficients $K$, $L$ and $F$ satisfy the relation, $$2\,L+A^2\,K=-2\,F\,.$$ We conclude that the linear in $A$ contribution to the l.h.s. of the equation (\ref{e11}) given initially by (\ref{e15}) becomes,
\eq{\label{e16}
&\varepsilon^{abc}\,A^e\,\left[2\,F^\prime-\chi\,F-L\right]+\\
&\varepsilon^{abe}\,A^c\,\left[\left(1+A^2\,\chi\right)\,G+2\,F+2\,(\chi+A^2\,\chi^\prime)\,F\right]+\\
&\varepsilon^{abm}\,A_m\,\delta^{ce}\,\left[G-\chi\,F-L\right]\,.
}

Again we set to zero the coefficients in (\ref{e16}) and obtain the relations
\eq{\label{e18}
 2\,F^\prime=\chi\,F+L\,.
}
\eq{\label{e17a}
\left(1+A^2\,\chi\right)\,G+2\,F+2\,(\chi+A^2\,\chi^\prime)\,F=0\,,}
and
\eq{\label{e17}
 G=\chi\,F+L\,,
}
The solution of the equation (\ref{e18}) with the initial condition, $F(0)=1$, is
\eq{\label{sol}
F(t)=\frac{\sin\sqrt{t}\cos\sqrt{t}}{\sqrt{t}}\,.
}
The relation (\ref{e17}) defines the function $G$ in terms of previously found ones $\chi$ and $F$.
The equation (\ref{e17a})
is satisfied as a consequence of the relation (\ref{e17}) and the differential equation (\ref{ode}). The same happens, for exemple, with the equation, \eq{\label{e40}L^\prime-\chi\,L=0\,,} resulting from the quadratic contribution (\ref{e20}). To show (\ref{e40}) one needs (\ref{ode}), (\ref{e14}) and (\ref{e18}). The careful check shows that the rest of the coefficients also vanishes. 
We stress that in order to the eq. (\ref{e11}) hold the function $\chi(t)$ cannot be arbitrary, but necessarily the one which guarantees the condition (\ref{an2}), i.e., $[\delta_{f},\delta_g] A
                      =\delta_{\{f,g\} }A \, .$

\subsubsection*{Definition of $R$-term}

Now using (\ref{an1}) and (\ref{e3}) in (\ref{s9}) we obtain after simplification,
\eq{
\label{e4}
R^{abc}\left(A\right)=&\left(\varepsilon^{abc}+\delta^{ab}\,A^c-\delta^{ac}\,A^b\right)\,S\left(A^2\right)\\
&+\left(\varepsilon^{abm}A_m\,A^c-\varepsilon^{acm}A_m\,A^b\right)\,T\left(A^2\right)\,,
}
where
\eq{\label{e23}
S=\frac{F}{2\,\left(1+A^2\,\chi\right)}\,,\qquad\mbox{and}\qquad  T=\frac{\chi\,F}{2\,\left(1+A^2\,\chi\right)}\,.
}  

\subsubsection*{Comparison to the lower order brackets}

As a consistency check let us calculate the first order contributions to the equations of motion. Since
\eq{\label{e30}
L=M=-\frac{\sin^2\sqrt{t}}{t}\,,\qquad S=V=\frac{\sin^2\sqrt{t}}{2t}\,,
}
one finds, $L(0)=M(0)=-1$. Then $N(0)=2F(0)=2$, and $S(0)=1/2$, so the first order contribution is given by
\eq{\label{e31}
2A_b\partial_aA_b- A_a\partial_bA_b-A_b\partial_bA_a+\frac12\,\varepsilon^{abc}\{A_b,A_c\}\,,
}   
which is in the perfect agreement with (\ref{nceom}). Now, 
\eq{\label{e32}
F^\prime(0)=  -\frac23\,,\qquad G(0)=-\frac43\,,\qquad V(0)=\frac12\,,
} 
which results in
\eq{\label{e33}
 -\frac23\,\varepsilon^{abc}A^2\partial_bA_c-\frac{4}{3}\,\varepsilon^{abm}A_m\,A^c\,\partial_bA_c+\{A_aA^2\} \,.
 }     
The term with the Poisson bracket is exactly the same as in (\ref{nceom}), but the coefficients at the first two terms are different. However, adding to the (\ref{e33}) the algebraic identity (\ref{E2}) from the Appendix multiplied by the factor $-2$,
$$
-2\,\varepsilon^{abc}\,A^2+2\,\varepsilon^{bcm}A_m\,A^a-2\,\varepsilon^{acm}A_m\,A^b+2\,\varepsilon^{abm}A_m\,A^c\equiv0\,,
$$ we arrive exactly to the equation (\ref{nceom}).

\subsection{Non-commutative field strength}

So far working in $3d$ we have constructed a vector ${\cal F}^a(A)$, which transforms covariantly under the NC gauge transformations (\ref{an1}), i.e., $ \delta_{f}  {\cal F}^a = \{f,{\cal F}^a\}$. The commutative limit of this vector gives the l.h.s. of the abelian Chern-Simons equations, $\lim_{\theta\to0}{\cal F}^a(A)=\varepsilon^{abc}\,\partial_bA_c$. Now, using the Levi-Ciivita epsilon we may define the tensor
\eq{\label{Fab}
{\cal F}^{ab}:=\varepsilon^{abc}  {\cal F}_c=P^{abcd}\left(A\right)\,\partial_c A_d+R^{abcd}\left(A\right)\,\left\{A_c,A_d\right\}\,,
}
which also transforms covariantly, $ \delta_{f}  {\cal F}^{ab} = \{f,{\cal F}^{ab}\}$, and reproduces in the commutative limit the abelian field strength,
\eq{
\lim_{\theta\to0}\,{\cal F}_{ab} =\partial_a A_b-\partial_bA_a\,.
}
We call this tensor the non-commutative field strength. Its explicit form is given by,
\eq{\label{ncfs}
{\cal F}^{ab}=&\,\,\partial^a\left(A^b\,F \big({\theta^2A^2}\big)\right)-\partial^b\left(A^a\,F \big({\theta^2A^2}\big)\right)\\
 &+\theta\,\varepsilon^{abc}\partial^d\big(A_c\,A_d\,L \big({\theta^2A^2}\big)\big)+\theta\, F \big({\theta^2A^2}\big)\varepsilon^{abc}\,\partial_c A^2\\
 &- \frac12 \{A^a L \left({\theta^2A^2}\big), A^b\right\}- \frac12 \{A^a, A^b L \big({\theta^2A^2}\big)\}-\frac{\theta}{2} L \big({\theta^2A^2}\big)\varepsilon^{abc}\,\{A_c, A^2\}\,,
}
where
\begin{equation*}
F \big({\theta^2A^2}\big)=\frac{\sin \big(2\sqrt{\theta^2A^2}\big)}{2\sqrt{\theta^2A^2}}\,,\qquad\mbox{and}\qquad L \big({\theta^2A^2}\big)=-\frac{\sin^2 \sqrt{\theta^2A^2}}{\theta^2A^2}\,.
\end{equation*}
Thus, just like in the commutative case, the non-commutative equations of motion, ${\cal F}^a(A)=0$, are satisfied if the non-commutative field strength vanishes everywhere, ${\cal F}^{ab}(A)=0$.

\subsection{Action principle}

To check whether exist an action $S$ yielding the field equations, ${\cal F}_a=0$, we use the criterium of second variational derivatives. If the equations are Lagrangian, i.e.,  $ {\cal F}_a= \delta {S}/ \delta A^a$, then
\eq{\label{lag}
\frac{\delta {\cal F}_a}{\delta A^b}=\frac{ \delta{\cal F}_b}{\delta A^a}\,.
}
One may easily check that, for 
$
{\mathcal  F}^a=P^{abc}\left(A\right)\,\partial_b A_c+R^{abc}\left(A\right)\,\left\{A_b,A_c\right\}\,,
$
 this condition does not hold.
 In particular, because $P^{abc}$ given by (\ref{e3}) is not an antisymmetric in $a$ and $c$.

On the other hand, there is a known result about the rigidity of the Chern-Simons action \cite{BH}, meaning essentially that up to the field redefinition any consistent deformation of the Chern-Simons action is proportional to the trivial one. Thus the absence of the action principle for the equation (\ref{e2}) means that possibly we obtained here some non-trivial deformation of the Chern-Simons theory. 

\subsection{Summary of the results}

Let us summarise the main results of the Section 5. Consider the three dimensional space endowed with the Poisson bracket,
\eq{\label{PB}
\{x^i, x^j\}=2\,\theta\,\varepsilon^{ijk}x_k\,.
}
In the subsection $5.1$ we have constructed the non-commutative deformation of the abelian gauge transformation, given by,
\eq{\label{gt}
 \delta_{f}  A_a=\partial_af+\{A_a,f\}+\theta\,\varepsilon^{abc}A_b\partial_cf+\theta^2\,\left(\partial_af A^2-\partial_bfA^bA_a\right)\chi\left(\theta^2A^2\right)\,,
}
where
\eq{
\chi(t)=\frac1t\,\left(\sqrt{t}\cot\sqrt{t}-1\right)\,,\qquad \chi(0)=-\frac13\,,
} 
which close the algebra
\eq{
[\delta_{f},\delta_g] A_a
                      =\delta_{\{f,g\} }A_a \,.
                      }    

The field equations covariant under the gauge transformation (\ref{gt}) and reproducing the abelian Chern-Simons equations in the commutative limit were obtained in the subsection $5.2$. An explicit form is,
\eq{
\label{s1}
   {\mathcal  F}^a:=P^{abc}\left(A\right)\,\partial_b A_c+R^{abc}\left(A\right)\,\left\{A_b,A_c\right\}=0\,,
} 
where
 \eq{
\label{s2}
P^{abc}\left(A\right)=&\varepsilon^{abc}\,F\left(\theta^2A^2\right)+\theta^2\varepsilon^{abm}A_m\,A^c\,G\left(\theta^2A^2\right)+\theta^3A^a\,A^b\,A^c\,K\left(\theta^2A^2\right)\\&+\theta\,A^a\,\delta^{bc}\,L\left(\theta^2A^2\right)+\theta\,A^b\,\delta^{ac}\,M\left(\theta^2A^2\right)+\theta\,A^c\,\delta^{ab}\,N\left(\theta^2A^2\right)\,,
}
and
\eq{\label{s3}
R^{abc}\left(A\right)=&\varepsilon^{abc}\,S\left(\theta^2A^2\right)+\theta^2\left(\varepsilon^{abm}A_m\,A^c-\varepsilon^{acm}A_m\,A^b\right)\,T\left(\theta^2A^2\right)\\
&+\theta\,\left(\delta^{ab}\,A^c-\delta^{ac}\,A^b\right)\,S\left(\theta^2A^2\right)\,,
}
and the coefficient functions are determined in (\ref{HJ}-\ref{e21a}), (\ref{e17}), (\ref{sol}) and (\ref{e23}) as
\eq{\label{s4}
F(t)&=\frac{N(t)}{2}=\frac{\sin\sqrt{t}\cos\sqrt{t}}{\sqrt{t}}\,,\\
 G(t)&=\frac{2\,\sqrt{t}\,\cos 2\sqrt{t}-\sin 2\sqrt{t}}{2\,t\,\sqrt{t}}\,,\\
K(t)&=-4\,T(t)=-\frac{2\sin\sqrt{t}}{t^2}\left(\sqrt{t}\,\cos \sqrt{t}-\sin \sqrt{t}\right)\,,\\
L(t)&=M(t)=-2\,S(t)=-\frac{\sin^2\sqrt{t}}{t}\,.
}
The equations (\ref{s1}) are non-Lagrangian and as in the commutative case are equivalent to the requirement that the non-commutative field strength (\ref{ncfs}) should vanish everywhere. Further physical properties and applications will be discussed elsewhere.

\section{Conclusions}

To construct the L$_\infty$ structure with given initial terms one has to solve the L$_\infty$ relations, $\mathcal{J}_n=0$. The key observation we made in this work is that in each given order $n$ there is the consistency condition of the equation, $\mathcal{J}_n=0$, which is satisfied as a consequence of the previously solved L$_\infty$ relations, $\mathcal{J}_m=0$, $m< n$. Using this observation we were able to derive the recurrence relations for the construction of the L$_\infty$ algebra describing the NC deformation of the abelian Chern-Simons theory in the slowly varying field approximation. Using these recurrence relations we made a conjecture regarding the explicit all orders formula for the NC $su(2)$-like deformation of the gauge transformations $\delta_f A$ and the corresponding field equations, ${\mathcal  F}(A)=0$. The functional coefficients in the proposed ansatz were fixed from the closure conditions of the gauge algebra (\ref{an2}) and the requirement of the gauge covariance of the equations of motion (\ref{e1}) correspondingly.

We conclude that the problem formulated in the introduction regarding the existence of the solution to the L$_\infty$ bootstrap program has a positive answer. Moreover we were able to find an explicit example of such a solution. Thus we can see that L$_\infty$ algebra is not only a correct mathematical framework to deal with the deformations but also is a powerful tool for the construction of these deformations. 

\subsection*{Acknowledgments}

I am grateful to Ralph Blumenhagen for fruitful discussions and to the Max Planck Institute for Physics for hospitality. I am also appreciative to the anonymous referees for the careful reading of the manuscript and valuable suggestions and remarks. This work was supported by the Grant 305372/2016-5 from the Conselho Nacional de Pesquisa (CNPq, Brazil) and the
Grant 2019/11475-6 from the Funda\c{c}\~{a}o de Amparo \'a Pesquisa do
Estado de S\~ao Paulo (FAPESP, Brazil). 

\section{Appendix: Important algebraic relations}

Since we are in $3d$, for any vector $A^e$ one may check that,
\eq{\label{E1}
\varepsilon^{abc}\,A^e-\varepsilon^{bce}\,A^a+\varepsilon^{cea}\,A^b-\varepsilon^{eab}\,A^c\equiv0\,.
}
The latter reflects the fact that in $3d$ any totally antisymmetric tensor of rank four vanishes.
Contracting the above identity with $A_e$ we arrive at,
\eq{\label{E2}
\varepsilon^{abc}\,A^2-\varepsilon^{bcm}A_m\,A^a+\varepsilon^{acm}A_m\,A^b-\varepsilon^{abm}A_m\,A^c\equiv0\,.
}
Taking the derivative of (\ref{E2}) with respect to $A_e$ one finds,
\eq{\label{E3}
&2\,\varepsilon^{abc}\,A^e-\varepsilon^{abe}\,A^c-\varepsilon^{abm}A_m\,\delta^{ce}+\\
&\varepsilon^{ace}\,A^b+\varepsilon^{acm}A_m\,\delta^{be}-\varepsilon^{bce}\,A^a-\varepsilon^{bcm}A_m\,\delta^{ae}=0\,.
}
Now using (\ref{E1}) in (\ref{E3}) we end up with
\eq{\label{E4}
\varepsilon^{abc}\,A^e-\varepsilon^{abm}A_m\,\delta^{ce}+\varepsilon^{acm}A_m\,\delta^{be}-\varepsilon^{bcm}A_m\,\delta^{ae}\equiv0\,.
}

One more identity we need is
\eq{\label{E5}
\varepsilon^{acm}A_m\varepsilon^{ben}A_n&=\left(\delta^{ab}\,\delta^{ce}-\delta^{ae}\,\delta^{bc}\right)\,A^2\\
&+\delta^{bc}\,A^a\,A^e-\delta^{ce}\,A^a\,A^b-\delta^{ab}\,A^c\,A^e+\delta^{ae}\,A^b\,A^c\,.
}
It can be obtained from (\ref{E2}) contracting it with $\varepsilon^{cen}$ and then renaming the indices.


\begin{thebibliography}{99}

\bibitem{Doplicher}
  S.~Doplicher, K.~Fredenhagen and J.~E.~Roberts,
  ``The Quantum structure of space-time at the Planck scale and quantum fields,''
  Commun.\ Math.\ Phys.\  {\bf 172} (1995) 187

 \bibitem{SW}
  N.~Seiberg and E.~Witten,
  ``String theory and noncommutative geometry,''
  JHEP {\bf 9909} (1999) 032
  
  \bibitem{BFFLS}
  F.~Bayen, M.~Flato, C.~Fronsdal, A.~Lichnerowicz and D.~Sternheimer,
  ``Quantum Mechanics as a Deformation of Classical Mechanics,''
  Lett.\ Math.\ Phys.\  {\bf 1} (1977) 521.
  
\bibitem{HS}
  C.~Hull and R.~J.~Szabo,
  ``Noncommutative gauge theories on D-branes in non-geometric backgrounds,''
  arXiv:1903.04947 [hep-th].
  
\bibitem{Wess}
  J.~Madore, S.~Schraml, P.~Schupp and J.~Wess,
  ``Gauge theory on noncommutative spaces,''
  Eur.\ Phys.\ J.\ C {\bf 16} (2000) 161

\bibitem{Dimitrijevic}
  M.~Dimitrijevic, F.~Meyer, L.~Moller and J.~Wess,
  ``Gauge theories on the kappa Minkowski space-time,''
  Eur.\ Phys.\ J.\ C {\bf 36} (2004) 117
  
\bibitem{Vassilevich}
  D.~V.~Vassilevich,
  ``Twist to close,''
  Mod.\ Phys.\ Lett.\ A {\bf 21} (2006) 1279
  
  \bibitem{Szabo}
  R.~J.~Szabo,
  ``Symmetry, gravity and noncommutativity,''
  Class.\ Quant.\ Grav.\  {\bf 23} (2006) R199
  
  \bibitem{BBKL}
  R.~Blumenhagen, I.~Brunner, V.~Kupriyanov and D.~L\"ust,
  ``Bootstrapping non-commutative gauge theories from L$_\infty$ algebras,''
  JHEP {\bf 1805} (2018) 097
  
 \bibitem{Zwiebach}
  B.~Zwiebach,
  ``Closed string field theory: Quantum action and the B-V master equation,''
  Nucl.\ Phys.\ B {\bf 390} (1993) 33
  
  \bibitem{HZ}
  O.~Hohm and B.~Zwiebach,
  ``$L_{\infty}$ Algebras and Field Theory,''
  Fortsch.\ Phys.\  {\bf 65} (2017) no.3-4,  1700014
  
\bibitem{Saemann}
  B.~Jurco, L.~Raspollini, C.~Saemann and M.~Wolf,
  ``$L_\infty$-Algebras of Classical Field Theories and the Batalin-Vilkovisky Formalism,''
  arXiv:1809.09899 [hep-th].
  
\bibitem{Stasheff1}
  T.~Lada and J.~Stasheff,
  ``Introduction to SH Lie algebras for physicists,''
  Int.\ J.\ Theor.\ Phys.\  {\bf 32} (1993) 1087
  
\bibitem{Stasheff2}
  J.~Stasheff,
  ``$L_\infty$ and $A_\infty$ structures: then and now,''
  arXiv:1809.02526 [math.QA].
  
\bibitem{Kontsevich}
  M.~Kontsevich,
  ``Deformation quantization of Poisson manifolds. 1.,''
  Lett.\ Math.\ Phys.\  {\bf 66} (2003) 157
   
\bibitem{kup-durham}
  V.~G.~Kupriyanov,
  ``$L_\infty$-Bootstrap Approach to Non-Commutative Gauge Theories,'' Fortsch.\ Phys.\ 
  doi:10.1002/prop.201910010
  arXiv:1903.02867 [hep-th].
  
\bibitem{Berends:1984rq}
  F.~A.~Berends, G.~J.~H.~Burgers and H.~van Dam,
  ``On the Theoretical Problems in Constructing Interactions Involving Higher Spin Massless Particles,''
  Nucl.\ Phys.\ B {\bf 260} (1985) 295
  
\bibitem{Fulp:2002kk}
  R.~Fulp, T.~Lada and J.~Stasheff,
  ``sh-Lie algebras induced by gauge transformations,''
  Commun.\ Math.\ Phys.\  {\bf 231} (2002) 25
  
\bibitem{BBKT}
  R.~Blumenhagen, M.~Brinkmann, V.~Kupriyanov and M.~Traube,
  ``On the Uniqueness of L$_\infty$ bootstrap: Quasi-isomorphisms are Seiberg-Witten Maps,''
  J.\ Math.\ Phys.\  {\bf 59} (2018) no.12,  123505
  
   \bibitem{KV}
V.~G.~Kupriyanov and D.~V.~Vassilevich,
  ``Star products made (somewhat) easier,''
  Eur.\ Phys.\ J.\ C {\bf 58} (2008) 627-637
  
\bibitem{Kupriyanov:2018yaj}
  V.~G.~Kupriyanov,
  ``Recurrence relations for symplectic realization of (quasi)-Poisson structures,''
  J.\ Phys.\ A {\bf 52} (2019) no.22,  225204
  
 \bibitem{LMPS} 
D.~Luest, E.~Malek, E.~Plauschinn and M.~Syvari,
  ``Open-String Non-Associativity in an R-flux Background,''
  arXiv:1903.05581 [hep-th].
  
 \bibitem{Hammou:2001cc}
  A.~B.~Hammou, M.~Lagraa and M.~M.~Sheikh-Jabbari,
  ``Coherent state induced star product on $R^3_\lambda$ and the fuzzy sphere,''
  Phys.\ Rev.\ D {\bf 66} (2002) 025025
  
\bibitem{GraciaBondia:2001ct}
  J.~M.~Gracia-Bondia, F.~Lizzi, G.~Marmo and P.~Vitale,
  ``Infinitely many star products to play with,''
  JHEP {\bf 0204} (2002) 026
  [hep-th/0112092].
  
 \bibitem{Vitale:2012dz}
  P.~Vitale and J.~C.~Wallet,
  ``Noncommutative field theories on $R^3_\lambda$: Toward UV/IR mixing freedom,''
  JHEP {\bf 1304} (2013) 115
  
\bibitem{Galikova:2013zca}
  V.~Galikova, S.~Kovacik and P.~Presnajder,
  ``Laplace-Runge-Lenz vector in quantum mechanics in noncommutative space,''
  J.\ Math.\ Phys.\  {\bf 54} (2013) 122106
  
\bibitem{Kupriyanov:2012nb}
  V.~G.~Kupriyanov,
  ``A hydrogen atom on curved noncommutative space,''
  J.\ Phys.\ A {\bf 46} (2013) 245303
    
\bibitem{Kupriyanov:2015uxa}
  V.~G.~Kupriyanov and P.~Vitale,
  ``Noncommutative $ {\mathrm{\mathbb{R}}}^d $ via closed star product,''
  JHEP {\bf 1508} (2015) 024
  
\bibitem{BH}
  G.~Barnich and M.~Henneaux,
  ``Consistent couplings between fields with a gauge freedom and deformations of the master equation,''
  Phys.\ Lett.\ B {\bf 311} (1993) 123 
    

  
\end{thebibliography}
\end{document}